\documentclass[a4paper,12pt]{article}

\usepackage{amsmath,amssymb,amsfonts}
\usepackage[dvips]{graphicx}
\usepackage{epsfig}
\usepackage{bbding}
\usepackage{pifont}
\usepackage{wasysym}
\usepackage{tikz}
\usepackage{color}
\usepackage{amsmath}
\usepackage{amsfonts}
\usepackage{verbatim}
\usepackage{amssymb}
\usepackage{graphicx}
\usepackage{amsmath,amssymb,calc}
\usepackage{bbm}
\usepackage{setspace}
\usepackage{color}
\usepackage{amsmath}
\usepackage{amsfonts}
\usepackage{verbatim}
\usepackage{amssymb}
\usepackage{graphicx}
\usepackage{array}
\usepackage{epstopdf}

\input{my_opt.sty}

\usepackage{amsmath,amssymb,amsfonts,graphicx}
\usepackage{epsfig}
\usepackage[left=2.4cm,top=3.3cm,right=2.4cm,bottom=3.3cm,bindingoffset=0cm]{geometry}

\usepackage{tabularx}

\def\be{\begin{equation}}
\def\ee{\end{equation}}
\def\bqa{\begin{eqnarray}}
\def\bea{\begin{eqnarray}}
\def\eea{\end{eqnarray}}
\def\beq{\begin{equation}}
\def\eeq{\end{equation}}

\def\brc{\langle}
\def\ckt{\rangle}

\def\Z{\mathbb Z}
\def\1{\mathbbm{1}}

\numberwithin{equation}{section}

\def\be{\begin{equation}}
\def\ee{\end{equation}}
\def\bqa{\begin{eqnarray}}
\def\bea{\begin{eqnarray}}
\def\eea{\end{eqnarray}}
\def\beq{\begin{equation}}
\def\eeq{\end{equation}}

\def\tr{\rm tr}

\def\rmc{{\rm c}}

\def\brc{\langle}
\def\ckt{\rangle}

\def\Z{\mathbb Z}
\def\1{\mathbbm{1}}

\def\tk{\tilde k}

\begin{document}

\title{ \begin{flushright}\ \vskip -2.5cm {\footnotesize {IFUP-TH-2019}}\end{flushright}
\vskip 30pt
\bf{Gauging 1-form center symmetries in  simple $SU(N)$ gauge theories}
}
\vskip 20pt  
\author{  Stefano Bolognesi$^{(1,2)}$, 
 Kenichi Konishi$^{(1,2)}$,  Andrea Luzio$^{(3,2)}$  \\[13pt]
{\em \footnotesize
$^{(1)}$Department of Physics ``E. Fermi", University of Pisa}\\[-5pt]
{\em \footnotesize
Largo Pontecorvo, 3, Ed. C, 56127 Pisa, Italy}\\[2pt]
{\em \footnotesize
$^{(2)}$INFN, Sezione di Pisa,    
Largo Pontecorvo, 3, Ed. C, 56127 Pisa, Italy}\\[2pt]
{\em \footnotesize
$^{(3)}$Scuola Normale Superiore, Piazza dei Cavalieri, 7,  56126, Pisa, Italy}\\[12pt]
{ \footnotesize  stefano.bolognesi@unipi.it,  \,  kenichi.konishi@unipi.it, \,  andrea.luzio@sns.it  }  
}

\maketitle


\vskip 0pt

\begin{abstract}

Consequences of gauging exact ${\mathbbm Z}_k^C$ center symmetries in several simple $SU(N)$ gauge theories, where $k$ is a divisor of $N$, 
are investigated. 
Models discussed include:
the $SU(N)$ gauge theory with $N_f$ copies of Weyl fermions in self-adjoint single-column antisymmetric representation,
the well-discussed adjoint QCD,
QCD-like theories in which the quarks are in a two-index representation of $SU(N)$, 
and a chiral $SU(N)$ theory with  fermions in the symmetric as well as in anti-antisymmetric representations but without fundamentals. 
Mixed 't Hooft anomalies between the 1-form ${\mathbbm Z}_k^C$ symmetry and some
0-form (standard) discrete symmetry provide us with useful information about the infrared dynamics of the system.  
In some cases they give decisive indication to select only few possiblities for the infrared phase of the theory.

\end{abstract}
\newpage

\newpage

\tableofcontents

\section{Introduction}\label{sec:introduction}

Recently the ideas of generalized symmetries and higher-form gauging  have been applied 
to gain deeper insights on the infrared dynamics of some strongly-coupled $4$D gauge theories \cite{Seiberg}-\cite{BKTL}.   One of the key issues, which lead to many interesting consequences,  
 is the so-called  ${\mathbbm Z}_N^C$ center symmetry\footnote{Throughout the paper, a 1-form center symmetry will be indicated with a suffix $C$, to distinguish it from various 0-form (conventional) discrete chiral symmetries.}   in $SU(N)$ theories. Although the idea itself is a familiar one,\footnote{A precursor of the ideas is indeed  the center symmetry ${\mathbbm Z}_N^C$ in Euclidean  $SU(N)$ Yang-Mills theory at finite temperature, which acts on the Polyakov loop. The unbroken (or broken) center symmetry  by the VEV of the Polyakov loop, is a criterion of confinement (or de-confinement) phase \cite{Polyakov:1978vu}.}  it becomes more powerful when combined with the idea of ``gauging" 
such a discrete center symmetry \cite{Seiberg}-\cite{BKTL}  In some cases,  this leads to mixed  ([0-form]-[1-form]) 't Hooft anomalies;  they carry nontrivial information on possible infrared dynamics of the system.

The concept of gauging a discrete symmetry might sound a bit peculiar from the point of view of conventional idea of gauging a global flavor symmetry, i.e., that of taking the transformation parameters as functions of spacetime and turning it  to a local gauge symmetry.  Here the gauging of a 1-form discrete symmetry means identifying the field configurations 
related by it, and eliminating the associated redundancies. In the case of the ${\mathbbm Z}_N^C$ center symmetry in $SU(N)$ gauge theory, gauging it  effectively reduces the theory to $SU(N)/{\mathbbm Z}_N$ theory \cite{Seiberg}-\cite{BKTL}. We review below (Sec.~\ref{review}) 
how this procedure works for the case of a subgroup,  $\mathbb{Z}_k^{C} \subset  {\mathbbm Z}_N^C$ discrete center symmetry.

 The aim of the present paper is to apply  these new ideas to  several simple $SU(N)$ gauge theories, which possess exact ${\mathbbm Z}_k^C$ color center symmetries  ($k$ being a divisor of $N$),  and 
to examine the implications of gauging these discrete ${\mathbbm Z}_k^C$ center symmetries on their infrared dynamics.
 In some cases our discussion is a simple extension of (or comments on) the results already found in the literature; in most others the results presented here are new, to the best of our knowledge. 
Here we discuss the following models:  in Section \ref{sa} the $SU(N)$ gauge theory with $N_f$ copies of Weyl fermions in self-adjoint single column antisymmetric representation;   in Section  \ref{aqcd} 
the adjoint QCD discussed extensively in the literature;  in Section  \ref{two} QCD-like theories with quarks in two-index representations of $SU(N)$,  
and  in Section \ref{chiral} some chiral $SU(N)$ theories with  fermions in the symmetric as well as in anti-antisymmetric representations but without those in the  fundamental representation.  
We conclude in  Section  \ref{conc} with some general discussion.    Notes on Dynkin indices for some representations in $SU(N)$ group can be found in Appendix~\ref{Dynkin}.

\section{Gauging a discrete 1-form symmetry \label{review}}

As the gauging of a discrete center symmetry and the calculation of anomalies 
under such gauging  are the basic tools of this paper and will be used repeatedly below, let us briefly review the procedure here. 
The procedure was  formulated in \cite{GKSW} and used in \cite{GKKS} for $SU(N)$ Yang-Mills theory at $\theta=\pi$,  based on and building upon some earlier results \cite{Seiberg}-\cite{AhaSeiTac},  
and then applied to other systems and further developed: see  \cite{TaniKiku}-\cite{AnbPop1}, and \cite{Popp}-\cite{Anber:2019nfu}.   The details and good reviews can be  found in  these references and will not be repeated here, except for a few basics reviewed below.

We recall that in order to gauge a  $\mathbb{Z}_k^{C}$ discrete center symmetry in an $SU(N)$ gauge theory  ($k$ being a divisor of $N$), 
 one introduces a pair of $U(1)$ 2-form and 1-form $\mathbb{Z}_k^{C}$  gauge fields $(B_c^{(2)},B_c^{(1)})$  satisfying the constraint   \cite{GKSW}
\be   {k}  B^{(2)}_\rmc =  d B^{(1)}_\rmc\;.\label{constr}
\ee
This constraint satisfies the invariance under the $U(1)$ 1-form gauge transformation, 
\beq
  B^{(2)}_\rmc \mapsto B^{(2)}_\rmc+d \lambda_\rmc,\qquad B^{(1)}_\rmc \mapsto B^{(1)}_\rmc+{k}\lambda_\rmc,     \label{Zn2}
\eeq
where
$\lambda_\rmc$ is  the 1-form gauge function, satisfying the quantized flux
\be  \frac{1}{2\pi}\int_{\Sigma_2}   d\lambda_\rmc  \in  {\mathbbm Z}  \;.
\ee 
The $SU(N)$ dynamical gauge field $a$ is embedded into a  $U(N)$ gauge field,
\be
\widetilde{a}=a+{1 \over k }B^{(1)}_\rmc, 
\ee
and one requires  invariance under $U(N)$ gauge transformation. The gauge field tensor $F(a)$ is replaced by 
\be   F(a) \to   {\tilde F}({\tilde a})  -  B^{(2)}_\rmc    \;.  
\ee
  This determines the way 
 these $\mathbb{Z}_k^{C}$ gauge fields are coupled  to the standard gauge fields  $a$;    the matter  fields must also be 
 coupled to the $U(N)$  gauge fields, such that the 
 1-form gauge invariance, (\ref{Zn2}), is  satisfied.  For a Weyl fermion $\psi$ this is achieved by  writing the  fermion   kinetic term as 
\be     {\bar \psi}   \, \gamma^{\mu}    ( \partial  +  R({\tilde a}) -  \frac {n(R)}{k}  B_c^{(1)})_{\mu}  P_L   \, \psi  \;, 
\ee
with $R({\tilde a}) $ appropriate for the representation to which $\psi$ belongs, and $n(R)$ is the $N$-ality of the representation  $R$.   $P_L$ is the projection operator on the left-handed component of the 
Dirac spinor.  This whole  procedure effectively eliminates the $\mathbb{Z}_k^{C}$ redundancy and defines a $SU(N)/\mathbb{Z}_k$ theory. 

Also, in order to study the anomaly of $U_{\psi}(1)$
symmetry  (or  of a discrete subgroup of it), $\psi  \to   e^{ i \alpha}  \psi$,  we introduce an external $U_{\psi}(1)$ gauge field $A_{\psi}$,
and couple it to the fermion as 
\be     {\bar \psi}   \, \gamma^{\mu}    ( \partial  +  R({\tilde a})  -  \frac {n(R)}{k}  B_c^{(1)}   +  A_{\psi}  )_{\mu}  P_L   \, \psi  \;.
\ee

It is easy now to compute the anomalies following the standard  Stora-Zumino descent  procedure \cite{Zumino,StoraZumino}, 
see also  \cite{Tanizaki}.  For simplicity   we write the expressions for a single fermion and for an Abelian symmetry, but these can be readily 
generalized.   A good recent review of this renowned constructions can be found in \cite{2groups}.
According to this procedure,  the anomaly can be evaluated starting from a $6$D (six-dimensional) Abelian anomaly \cite{Zumino,StoraZumino}\footnote{$D(R)$  is the value of the symmetric trace of the product of three generators normalized to the one evaluated in the fundamental representation;  $T(R)$ is the Dynkin index
of the representation $R$, see Appendix~\ref{Dynkin}.   Throughout, the simplified differential form notation is used, e.g.,
$F^2 \equiv  F \wedge F =   \frac{1}{2}  F^{\mu \nu} F^{\rho \sigma} dx_{\mu} dx_{\nu} dx_{\rho} dx_{\sigma}= \frac{1}{2}
\epsilon_{\mu \nu \rho \sigma}   F^{\mu \nu} F^{\rho \sigma}  d^4x $, etc. }  
\bea   &&
 {1
 \over 24  \pi^2}{\tr}_{R} \big( {\tilde F}-B^{(2)}_c    +  d A_{\psi}  \big)^3\nonumber \\
&&  =   {  D(R) 
 \over 24  \pi^2}{\tr} \big( {\tilde F}-B^{(2)}_c )^3  +  {  2\, T(R)    \over 8\pi^2}    {\tr} \big( {\tilde F}-B^{(2)}_c\big)^2 \wedge dA_{\psi} + \ldots   \label{from}
\eea
Let us recall that,   in the standard quantization (i.e., in the absence of the 1-form discrete symmetry gauging), 
\be      B^{(1)}_\rmc, B^{(2)}_\rmc \to 0\;,    \qquad  {\tilde F}({\tilde a}) \to   F(a)\;,   
\ee
and  the above reduces to 
\be 
  {  D(R) 
 \over 24  \pi^2}{\tr}   F^3 +      { 2  T(R)    \over 8\pi^2}    {\tr}   F^2 \wedge dA_{\psi} + \ldots 
 \label{picking}
\ee
By using the identity  \cite{StoraZumino,Zumino} 
\be   
 {\tr}  F^3  =  d\, \{  {\tr}   (   a (d a)^2 + \frac{3}{5}  (a)^5 + \frac{3}{2} a^3 d a)   \}\;
\ee
(also   $   {\tr}  F^2  =  d \, \{  {\tr}   (  a d a + \frac{2}{3}  a^3 )\}$)
the first term leads to the $SU(N)$ gauge anomalies.  
  The
second term gives the boundary term 
\be  \frac{ 2 T(R) }{ 8\pi^2 }  \int_{\Sigma_5}   {\tr}  F^2 \wedge   A_{\psi} 
\ee
which,  after variations
\be      A_{\psi}  \equiv d A_{\psi}^{(0)}\;, \qquad       A_{\psi}^{(0)} \to A_{\psi}^{(0)}+  \delta \alpha 
\ee
yields, by anomaly inflow,   the well-known $4D$ anomaly,  
\be  \delta  S_ { \delta A_{\psi}^{(0)} }  =      {2 T(R)  \over 8\pi^2}   \int   {\tr}  F^2   \,    \delta \alpha    =    2  \, T(R)  \,     {\mathbbm Z} \,   \delta   \alpha\;,  
\ee
where $ {\mathbbm Z} $ represents  the integer  instanton number, leading to the well-known result that the discrete subgroup 
\be     {\mathbbm Z}_{2  T(R)} \subset U_{\psi}(1)
\ee
remains unbroken by  instantons. 

With the 1-form gauging in place, i.e., with $(B_c^{(2)}, B_c^{(1)}$) fields present in Eq.~(\ref{from}),  $ U_{\psi}(1)$ symmetry could be further broken
to a smaller discrete subgroup,  due to the replacement, 
\be     {\tr}   F^2  \to     {\tr} \big( {\tilde F}-B^{(2)}_c\big)^2\;.
\ee
In fact \footnote{Observe that $B^{(2)}_c$  is Abelian, $\propto {\mathbbm 1}_N$, and that  $  {\tr}   {\tilde F} = N    \, B^{(2)}_c$.},
\be  \frac{1}{8\pi^2}  \int_{\Sigma_4}     {\tr}     \big( {\tilde F}-B^{(2)}_c\big)^2 =  \frac{1}{8\pi^2}   \int_{\Sigma_4}    \{   {\tr}   {\tilde F}^2 -  N    (B^{(2)}_c)^2 \} \;:
\ee
the first term is an integer \footnote{The combination 
\[ \frac{1}{8\pi^2}   \int_{\Sigma_4}    \{   {\tr}   {\tilde F}^2 -  {\tr}   {\tilde F} \wedge  {\tr}   {\tilde F} \}  \]
is the second Chern number of $U(N)$ and is an integer.  The second term of the above is also an integer as $(\frac {N}{k})^2$ is. 
}; the second term is 
\be   -     \frac{N}{8\pi^2}    \int_{\Sigma_4} \big( B^{(2)}_c\big)^2  =   -     \frac{N}{8\pi^2  k^2}    \int_{\Sigma_4} \big( dB^{(1)}_c  \wedge dB^{(1)}_c  \big )     = \frac{N}{k^2}\, {\mathbbm Z}\;,
\ee
which is in general fractional.

\section{Models with self-adjoint  chiral fermions}
\label{sa}

We first consider  a class of  $SU(N)$ gauge theories ($N$ even) with  left-handed fermions in the $\frac N2$ fully antisymmetric representation. 
This representation is equivalent to its complex conjugate (as can be seen by acting on it with the epsilon tensor) and so does not contribute to the gauge anomaly. In these models there is a 1-form   $ {\mathbbm Z}_{\frac N2}^{C} $ center symmetry and we are particularly interested in understanding how this mixes in the 't Hooft anomalies with the other 0-form symmetries present.

\subsection{$SU(6)$  models  \label{sec:Yamaguchi}}

Let first examine in detail the case $N=6$ with  $N_f$ flavors of Weyl fermions   in the representation 
\be   {\underline{20}} \, = \, \yng(1,1,1)\ .
\ee
As will be seen below,  this  ($SU(6)$) is the simplest nontrivial case of interest.
  The  first coefficient of the beta function is
\be  
 b_0=  \frac{11 N -   6 N_f}{3}  = 22 -  2 N_f   \ ,
\ee
so up to $N_f=10$ flavors are allowed for the theory to be asymptotically free.  
In all these models, as will be  explained in the following,  there is a $U(1)_{\psi}$ global symmetry broken by the usual ABJ anomaly and instantons  to a global discrete  ${\mathbbm Z}_{6 N_f}^{\psi}$ which is then further broken by the 1-form gauging to  ${\mathbbm Z}_{2 N_f}^{\psi}$.
Note that  the latter breaking should be understood in the sense of a mixed  't Hooft anomaly:  there is an obstruction to gauging such a 
$ {\mathbbm Z}_{\frac N2}^{C} $ discrete center symmetry, while trying to maintain the global   ${\mathbbm Z}_{6 N_f}^{\psi}$ symmetry.
.

%

\subsubsection{$N_f = 1$}

Let us further restrict ourselves to   $SU(6)$ theory with  a {\it single} left-handed fermion in the representation,  ${\underline {20}}$.
This model was considered  recently in \cite{Yamaguchi}.   A good part of the analysis below indeed overlaps with \cite{Yamaguchi}; nevertheless,
we discuss this simplest model with certain care, in order to fix the ideas, to recall the basic techniques and notations, and to discuss physic questions involved. 

There are no continuous nonanomalous  symmetries in this model. 
There is an anomalous $U(1)_{\psi}$ symmetry whose nonanomalous subgroup is the  ${\mathbbm Z}_6^{\psi}$ symmetry given by
\be   
{\mathbbm Z}_6^{\psi}\,:\quad  \psi \to    e^{\frac{2\pi i}{6}  j} \psi \ ,   \qquad  j=1,2,\dots ,6  \ .   \label{flavorZ6}
\ee
The system possesses also an exact center symmetry which acts on Wilson loops as 
\be     
 {\mathbbm Z}_3^{C}  \,: \quad  e^{ i \oint A } \to         e^{\frac{2\pi i}{6} k}  e^{ i \oint A }   \ ,      \qquad  k=2,4,6 \ ,
\ee
and which  does not act on $\psi$.

This is an example of generalized symmetries (in this case, a $1$-form symmetry), which have received a 
considerable (renewed) attention in the last several years.  In particular, the central idea is that of gauging a discrete symmetry (such as 
${\mathbbm Z}_3^{C}$ here), i.e.,  that of  identifying field configurations related by those symmetries, and effectively modifying the 
path-integral sum over them.  If a center ${\mathbbm Z}_k^{C}$  symmetry is gauged,   $SU(N)$ gauge theory is replaced by
$SU(N)/{\mathbbm Z}_k$ theory.  The basic aspects of such a procedure were reviewed in Sec.~\ref{review}.

Let us now apply this method to our simple  $SU(6)$ toy model,  to study the fate of the unbroken $ {\mathbbm Z}_6^{\psi}$ symmetry
(\ref{flavorZ6}),  
  in the presence of  the  
$ {\mathbbm Z}_3^{C}$  gauge fields. 
The Abelian $6D$ anomaly takes the form  \footnote{
The trace  $  {\tr} $  without specification of the representation means that it is taken on the fundamental, $\underline {N}$. }
\bea   && \frac{1}
 {24  \pi^2}\, {\tr}_{ {\underline{20}} } \big( {\tilde F} - B^{(2)}_c    -   d A_{\psi}  \big)^3\nonumber \\
&&  =    { 6 \over 8\pi^2}    {\tr} \big( {\tilde F}-B^{(2)}_c\big)^2 \wedge dA_{\psi} + \ldots   \nonumber \\
 &&  =    { 6 \over 8\pi^2}   {\tr}  {\tilde F}^2  \wedge dA_{\psi}  -  { 6 N \over 8\pi^2}   (B_c^{(2)})^2  \wedge dA_{\psi}   + \ldots   \label{formula1}
\eea
where  $ {\mathbbm Z}_3^{C}$ gauge fields satisfy
\be   {3}  B^{(2)}_\rmc =  d B^{(1)}_\rmc\;,  \label{constr11}
\ee
invariant  under the $U(1)$ 1-form gauge transformation, 
\beq
  B^{(2)}_\rmc \mapsto B^{(2)}_\rmc+d \lambda_\rmc\,,\quad B^{(1)}_\rmc \mapsto B^{(1)}_\rmc+{3}\lambda_\rmc\,,     \label{Z311}
\eeq
\be  \frac{1}{2\pi}\int_{\Sigma_2}   d\lambda_\rmc  \in  {\mathbbm Z}  \;.\label{quantiz1}
\ee 
The factor $6$ in (\ref{formula1}) is twice the Dynkin index of  $ {\underline{20}} $
 (see Appendix~\ref{Dynkin}).   $A_{\psi} $ is a $U(1)$ gauge field,  formally introduced to describe 
the ${\mathbbm Z}_6^{\psi}$ discrete symmetry transformations.

The first term in (\ref{formula1}) is clearly trivial, as
\be   {1 \over 8\pi^2}   \int   {\tr}  {\tilde F}^2  \in   {\mathbbm Z} \,, \qquad 
A_{\psi} =   d A_{\psi}^{(0)}\;, \quad       \delta   A_{\psi}^{(0)}   =  \frac{2 \pi  {\mathbbm Z}_6^{\psi}  }{6}    \;.
\ee
This corresponds to the standard gauge anomaly that breaks $ U(1)_{\psi} \longrightarrow {\mathbbm Z}_6^{\psi}$.

The second term  in (\ref{formula1}) shows that $\delta   A_{\psi}^{(0)} $ gets  multiplied by
\be    -   { 6  N    \over 8\pi^2}   \int    (B_c^{(2)})^2   =   -    6  N   \big(\frac{1}{3}\big)^2  {\mathbbm Z}  =   - 6   \,   \frac{2}{3}  {\mathbbm Z}\ . 
\ee
The crucial step used here is  the flux quantization of the $B_c^{(2)}$ field
\be     { 1 \over 8\pi^2}   \int    (B_c^{(2)})^2  =     \big(\frac{1}{3}\big)^2 \,  {\mathbbm Z} \;, 
\ee
which follows from    (\ref{constr11})-(\ref{quantiz1}) \footnote{The factor $3$ is replaced by $k$  in the case of  $ {\mathbbm Z}_k^{C}$  discrete center 
gauging considered below for other systems.}. 
The      global chiral ${\mathbbm Z}_6^{\psi}$  symmetry
\be  \delta   A_{\psi}^{(0)}   =    \frac{2\pi   \ell}{6}\ , \qquad \ell=1,2,\ldots, 6
\ee
is therefore reduced to  a ${\mathbbm Z}_2^{\psi}$ invariance obtained restricting to the elements $\ell=3,6$, 
\be
{\mathbbm Z}_6^{\psi} \longrightarrow {\mathbbm Z}_2^{\psi} \ .
\label{reduced}
\ee
This agrees with what was found by \cite{Yamaguchi}.
   This implies that a confining vacuum with mass gap, with no condensate formation and with unbroken ${\mathbbm Z}_6^{\psi}$,  is not consistent.
   
Strictly speaking, it is not quite correct to say that the flavor symmetry of the model is ${\mathbbm Z}_6^{\psi}$, Eq.~(\ref{flavorZ6}),  since 
${\mathbbm Z}_2^{\psi}  \subset  {\mathbbm Z}_6^{\psi}$   (i.e.,  $\psi \to -\psi$)   is shared with the color ${\mathbbm Z}_2^{C}  \subset  {\mathbbm Z}_6^{C}$. The correct symmetry is 
\be    \frac{  {\mathbbm Z}_6^{\psi}  } { {\mathbbm Z}_2  }    \sim    {\mathbbm Z}_3^{\psi}\;.  
\ee
However our conclusion is not modified:  (\ref{reduced}) is actually equivalent to 
\be
{\mathbbm Z}_3^{\psi} \longrightarrow {\mathbf 1} \,,
\label{reducedBis}
\ee
in the vacuum with unbroken ${\mathbbm Z}_2^{color-flavor}$.  This feature must be kept in mind in all our analysis below: the crucial point is that  in this paper we  gauge only a subgroup of discrete color center group, which does not act on the fermions \footnote{The situation is subtler if one tries to gauge the full color center group, see  \cite{BKTL}.}.

The breaking  ${\mathbbm Z}_6^{\psi} \to {\mathbbm Z}_2^{\psi}$  implies a threefold vacuum degeneracy,  if the system confines (with mass gap) and if  in 
IR there are no massless fermionic degrees of freedom on which  ${\mathbbm Z}_6^{\psi} / {\mathbbm Z}_2^{\psi}$ can act \footnote{Here, as in the rest of the paper, we do not consider the more ``exotic'' possibility that  discrete anomaly matching may be  achieved with a topological field theory or by a CFT  in the IR. }. A possible explanation naturally presents itself.  As the interactions become strong in the infrared, it is reasonable to assume that bifermion condensate
\be  \langle  \psi \psi  \rangle    \sim \Lambda^3  \ne 0 \label{thefact}
\ee
 forms.   As the field $\psi$ is in  ${\underline {20}}$ of the gauge group $SU(6)$,  a Lorentz invariant bifermion composite can be in one of the irreducible representations of $SU(6)$,  appearing in the decomposition
 \be  \yng(1,1,1) \otimes \yng(1,1,1) =  \yng(1,1,1,1,1,1)\oplus \yng(2,1,1,1,1)\oplus +\ldots\;.
 \ee
 The most natural candidate would be the first, ${\underline 1}$, but it can be readily verified that such a condensate vanishes by the Fermi-Dirac statistics. 
 Another possibility is that  $\psi \psi$ in the adjoint representation gets a VEV,  signaling a sort of dynamical Higgs mechanism  \cite{Raby,BKS,BK}.
 Even though such a condensate should necessarily be regarded as a gauge dependent expression of  some gauge invariant  VEV (see below), it
 unambiguously  signals \footnote{Note that the global symmetry group ${\mathbbm Z}_6^{\psi}$  commutes with the color $SU(6)$: there is no way a gauge transformation eliminates the nontrivial properties of the condensate under ${\mathbbm Z}_6^{\psi}$.}   the breaking of global, discrete   chiral symmetry as 
 \be   {\mathbbm Z}_6^{\psi}  \to    {\mathbbm Z}_2^{\psi},    \ee    
  with broken ${\mathbbm Z}^{\psi}_6 / {\mathbbm Z}^{\psi}_2$ acting on the degenerate vacua.  
 Four-fermion, gauge-invariant condensates such as  
\be     \langle  \psi \psi \psi \psi   \rangle    \ne 0 \;, \qquad {\rm or} \qquad     \langle  {\bar \psi } {\bar \psi}  \psi \psi   \rangle    \ne 0  \;,    \label{ginvariant}
\ee
might also form, first of which also breaks ${\mathbbm Z}^{\psi}_6$ in the same way.  
The condensate (\ref{thefact}) thus leads to threefold vacuum degeneracy, consistently with (\ref{reduced})  implied by the  $ {\mathbbm Z}^{\psi}_6 - {\mathbbm Z}_3^{C}$ mixed anomaly.

Let us pause briefly to make a few comments on dynamically induced Higgs phase. 
As in any system where the Higgs mechanism is at work, some  (elementary or composite) scalar field 
gets a nonvanishing, gauge noninvariant (and gauge dependent) vacuum expectation value  (VEV). 
In a weakly coupled Higgs type model, 
there is a potential having degenerate minima, and the vacuum, which necessarily breaks the gauge invariance, induces
the Higgs phase, with some gauge bosons becoming massive.  Also, in such a model, apparently gauge-dependent phenomena can be 
naturally re-interpreted 
in a gauge-invariant fashion \footnote{For instance, the Higgs VEV
of the form $\brc \phi \ckt   =    \left(\begin{array}{c} 0  \\  v / \sqrt{2} \end{array}\right)$ found in any textbook about the 
standard electroweak theory,   is just a gauge dependent way of describing a
minimum of the potential  $V( \phi^{\dagger} \phi )$,  so is the statement such as the left hand fermion  being equal to  $  \psi_L= \left(\begin{array}{c}\nu_L \\e_L\end{array}\right)$. A similar reinterpretation of the $W$ and $Z$ bosons is also straightforward.}.

 Here the situation is basically the same.    One is indeed assuming that an effective composite scalar  $\sim \psi \psi$ 
forms by strong interactions, which then 
condenses.  It corresponds to  the non-gauge-invariant VEV of a scalar field
in a potential model.  In contrast to a weakly coupled Higgs models,  however,  the effective scalar composite particle is still strongly coupled and is not
described by a simple potential.  Therefore,  a gauge-invariant rephrasing of the phenomenon may not be straightforward. 
Apart from this,  there is nothing unphysical  about assuming  gauge non-invariant bifermion condensate \footnote{In this respect we differ from the interpretation given in \cite{Yamaguchi}.   Indeed there is a long history of studies in strongly interacting chiral gauge theories based on such
ideas, starting from \cite{Raby}.  See also \cite{BKS,BK} and references cited therein. }: it is analogue of the Higgs VEV $\brc \phi \ckt $
in the standard electroweak theory.  

 As a final remark,  it may help to remember also that  the Higgs mechanism  itself was first discovered in the context of superconductivity (\cite{Nambu, Anderson}, see also \cite{Schwinger}):  the Cooper pair condenses due to the interactions between the electrons and the lattice phonons. The Cooper pair, having charge 2,  is not a gauge invariant object.  It is the first example in a physical theory of what we call dynamical Higgs mechanism, in the sense that the effective Higgs scalar (the Cooper pair) is a composite, gauge noninvariant field \footnote{This brief comment is meant only to remind the reader that there is nothing unusual in having a composite, gauge noninvariant field getting a VEV,  to break the gauge (and/or flavor)  symmetry of a given system.
 Of course, conventional superconductivity is not a good model for strongly interacting gauge theories as the ones we are interested in here.       
 }.

The infrared system depends also on the kind of bi-fermion $\psi\psi$  condensates which actually form.  The  ``MAC'' (most attractive channel) criterion \cite{Raby} suggests
 condensation of a  $\psi\psi$ composite scalar  in the adjoint representation. It is then possible that the infrared physics is described by full dynamical Abelianization \cite{BKS,BK}:  
the low-energy theory is an Abelian $U(1)^5$ theory.    
Although the infrared theory looks trivial,  the only massless infrared degrees of freedom being five types of non-interacting photons, the system might be richer actually. 
There is a remnant of the ${\mathbbm Z}_6$ symmetry of the UV theory, which is a threefold vacuum degeneracy.   Domain walls would exist which connect the three vacua, and on which nontrivial infrared  $3$D physics can appear  (we shall not elaborate on them here).

It is interesting to   check also the (conventional) $[\mathbbm{Z}^{\psi}_6]^3$ anomaly  matching constraint,  following  \cite{IbanezRoss,CsakiMura}. 
The matching condition for a  $[{\mathbbm Z}_N]^3$  discrete symmetry is 
\bea && A_{\rm IR}= A_{\rm UV} + m N  \ ,  \qquad \qquad \quad  {\rm for \  odd}\ N\ , \nonumber \\  &&  A_{\rm IR}=A_{\rm UV} + m N + \frac{nN^3}{8}\ ,    \quad    \ \   {\rm  for \  even}\  N \ , \label{dam} \eea
where $m,  n \in    {\mathbbm Z}$ \footnote{The difference between the even $N$ and  odd $N$ cases is due to the possibility that a single fermion $\psi$ with charge $N/2$ can get a Majorana mass term for even $N$ (which does not break the $Z_N$ symmetry). Such a fermion provides a $\frac{N^3}{8}$ contribution to the $[{\mathbbm Z}_N]^3$, therefore the anomaly matching conditions for massless fermions is weakened. For further details see \cite{CsakiMura}.}.    In our case $N=6$,  $\frac{N^3}{8}=27$ and it must be that 
\be  A_{\rm IR} -  A_{\rm UV}  =  0     \mod   3\;
\ee
if  ${\mathbbm Z}^{\psi}_6$ is to remain unbroken.  However, 
  $A_{\rm UV}\big([\mathbbm{Z}_6^{\psi}]^3\big)=20 =  2\, \mod 3  \ne  0$ in our system, where $20$ is the color multiplicity. Therefore a confining vacuum with no condensates with mass gap, and with unbroken  ${\mathbbm Z}^{\psi}_6$  ($A_{\rm IR}=0$),  would not be consistent.    On the other hand, the condensate formation (\ref{thefact}), a spontaneous breaking $\mathbb{Z}^{\psi}_6 \longrightarrow \mathbbm{Z}^{\psi}_2$ and associated threefold vacuum degeneracy, is perfectly consistent with the $[\mathbbm{Z}^{\psi}_2]^3$ anomaly  matching condition.
    Consideration of $\mathbbm{Z}^{\psi}_6\,[{\rm grav}]^2$ anomaly leads to the same conclusion.

 We find  thus that the consideration of the 1-form $ {\mathbbm Z}_3^{C} $  center symmetry gauging (\ref{reduced})
and that of the conventional 
   $[\mathbbm{Z}^{\psi}_6]^3$  or  $\mathbbm{Z}^{\psi}_6\,[{\rm grav}]^2$   anomaly  matching requirement, give
a consistent indication about the infrared dynamics of our system.

\subsubsection{$N_f = 2$}

We now move to discuss $SU(6)$ theory with more than one  Weyl fermions in ${\underline {20}}$.
For two flavors the global symmetry  is   
\be 
G_f  = SU(2)\times {\mathbbm Z}_{12}^{\psi} \ .   \label{isnot}
\ee 
As before there is a  $0$-form and $1$-form mixed  anomaly
\be     {\mathbbm Z}_{12}^{\psi}\,[ {\mathbbm Z}_{3}^{C}]^2\ ,\label{mixed}
\ee 
and the discrete chiral  symmetry is broken by the 1-form gauging  as 
\be    {\mathbbm Z}_{12}^{\psi}  \longrightarrow   {\mathbbm Z}_{4}^{\psi} \,.  \label{exactly}
\ee
 In this case the bi-fermion scalar condensate
\be     \langle  \psi^{[A} \psi^{B]}  \rangle   \ne 0  \ ,  \label{SU(2)cond}
\ee
can be formed which is gauge-invariant, and leaves  $SU(2)$ invariant. Let us assume that such a condensate indeed is formed.
The bi-fermion condensate (\ref{SU(2)cond})   breaks  the discrete chiral  symmetry
\be    {\mathbbm Z}_{12}^{\psi}  \longrightarrow  {\mathbbm Z}_{2}^{\psi} \ ,     \label{condbreaks}
\ee
implying a {\it six-fold} vacuum degeneracy.
The latter is stronger than (\ref{exactly}) but is consistent.   
The $SU(2)$ triangle anomaly vanishes so there are no associated matching constraints:   
neither massless baryons nor NG bosons are required, and expected to occur.
The Witten  $SU(2)$ anomaly is also matched  between the UV ($6$ doublets) and  the IR ($0$ doublet).

  Strictly speaking the symmetry of the $N_f=2$ system is not (\ref{isnot}), but
\be 
G_f  = \frac{  SU(2)\times {\mathbbm Z}_{12}^{\psi} }{  {\mathbbm Z}_2  \times  {\mathbbm Z}_2 } \sim    
\frac{  SU(2) }{  {\mathbbm Z}_2}   \times {\mathbbm Z}_{6}^{\psi}  \ .   \label{butis}
\ee 
where one of the factors in the denominators is due to the overlap with the ${\mathbbm Z}_2^{color}$, the other being a $ {\mathbbm Z}_2$ shared between 
$SU(2)$ and ${\mathbbm Z}_{12}^{\psi}$.    
A similar observation was made in the  $N_f=1$ case,  in Sec.(3.1.1). 
As  discussed in the previous case,  and as will be in all  other cases discussed below,  none of our conclusions is modified by this more
careful consideration of the symmetry group,  as  Eq.~(\ref{exactly}) and   Eq.~(\ref{condbreaks}) are  equivalent to 
\be    {\mathbbm Z}_{6}^{\psi}  \longrightarrow   {\mathbbm Z}_{2}^{\psi} \,,  \label{exactlyBis}
\ee
and
   \be    {\mathbbm Z}_{6}^{\psi}  \longrightarrow  {\mathbf 1} \ ,     \label{condbreaksBis}
\ee
respectively,    in the presence of unbroken $SU(2)$ and ${\mathbbm Z}_2^{color}$ (confinement).

As for the conventional discrete anomaly matching,  in the UV there is a discrete $0$-form 
\be    {\mathbbm Z}_{12}^{\psi}\,[SU(2)]^2 
\label{ca}
\ee
anomaly. This is due to the 
$SU(2)$ instanton effects, which gives the phase variation of the partition function,
\be   20 \, \frac{2\pi k}{12}  =  2\pi   \frac{5}{3}k\ , \qquad k=1,2,\ldots, 12\ . 
\ee
 breaking  the chiral symmetry as 
\be     {\mathbbm Z}_{12}^{\psi}  \longrightarrow   {\mathbbm Z}_{4}^{\psi} \;
\label{ofgb}
\ee
since only the  transformations with $k=3,6,9,12$ leave the partition function invariant.
One may wonder how such a breaking is described in the IR, where no massless fermions 
are present, and $SU(2)$ is unbroken. The answer is that the condensate  (\ref{SU(2)cond}) breaks the discrete chiral symmetry as (\ref{condbreaks}), which is stronger than  (\ref{ofgb}).

We consider also  the constraints following from the  $[{\mathbbm Z}_{12}^{\psi}]^3$ anomaly.  This time an unbroken ${\mathbbm Z}_{12}^{\psi}$  in the IR  requires
the anomaly matching (\ref{dam}) 
\be A_{\rm UV}= A_{\rm IR}+  12 \,m  + 6^3 \,n\ ,  \qquad   m, n \in {\mathbbm Z}\ ,
\ee
that is, 
\be     A_{\rm UV} - A_{\rm IR} = 0 \mod 12\ . 
\ee
The UV anomaly is 
\be    A_{\rm UV} =  4    \mod   12\ , 
\ee
therefore the $[{\mathbbm Z}_{12}^{\psi}]^3$ anomaly consideration is consistent with the assumption of the breaking 
$Z_{12}^{\psi} \longrightarrow \Z_4^{\psi}$   and with consequent   threefold  degeneracy of the vacuum. This is also
consistent with result of the 1-form gauging of the center  ${\mathbbm Z}_{3}^{C}$ symmetry and the mixed anomaly (\ref{mixed}).

To conclude, it is possible that actually  a bi-fermion gauge-invariant condensate (\ref{SU(2)cond}) forms, breaking the discrete symmetry, as $ {\mathbbm Z}_{12}^{\psi}  \longrightarrow   {\mathbbm Z}_{2}^{\psi}.$  It is however also possible that the bi-fermion condensate vanishes, and e.g., some four-fermion condensates are formed.  In that case, the discrete symmetry breaking pattern would coincide with what is implied by the conventional and mixed anomalies associated with the discrete ${\mathbbm Z}_{12}^{\psi}$ symmetry (\ref{ca}) and  (\ref{mixed}).

\subsubsection{$N_f = 3$}

For three fermions in ${\underline {20}}$  the symmetry is 
\be   G_f=    SU(3)\times {\mathbbm Z}_{18}^{\psi} \ .
\ee
The 1-form gauging of the  ${\mathbbm Z}_3^{C}$ center symmetry  yields 
the discrete  symmetry  breaking
\be      {\mathbbm Z}_{18}^{\psi}  \longrightarrow  {\mathbbm Z}_{6}^{\psi}\ . \label{weaker}
\ee
In this case, a gauge-invariant bi-fermion condensate, 
\be     \langle  \psi^{[A} \psi^{B]}  \rangle  \ , \qquad   A,B=1,2\ ,  \label{SU(3)cond}
\ee
 would break the continuous symmetry as
\be       SU(3) \longrightarrow     SU(2)\ . 
\ee
There are $8-3=5$  NG bosons, which saturate the anomalies of the spontaneously broken $SU(3)/SU(2)$ symmetries.  There are no triangle $SU(2)^3$
anomalies.  No massless  baryons are required and expected
to occur  in the infrared theory. 
The discrete ${\mathbbm Z}_{18}^{\psi}$ symmetry  would be broken by the condensate (\ref{SU(3)cond}) as  
\be      {\mathbbm Z}_{18}^{\psi} \longrightarrow    {\mathbbm Z}_{2}^{\psi}\ ,\label{condensatebr}
\ee
implying a nine-fold vacuum degeneracy.

Again, let us check   $  {\mathbbm Z}_{18}^{\psi}\,[SU(3)]^2$ (to be matched mod  $N=18$) and $[ {\mathbbm Z}_{18}^{\psi}]^3$ anomalies.
Actually,  since $SU(3)$ is broken to  $SU(2)$  by bi-fermion condensate, we shall study the $  {\mathbbm Z}_{18}^{\psi}\,[SU(2)]^2$.
As 
\be    A_{\rm UV} \big(  {\mathbbm Z}_{18}^{\psi} \,[SU(2)]^2 \big)= 20 =    2  \mod  18\ ,
\ee
the $ {\mathbbm Z}_{18}^{\psi}\,[SU(2)]^2$  anomaly matching is consistent with  what is implied by the  $\psi^2$ condensate formation (\ref{SU(3)cond}).

As for  $[ {\mathbbm Z}_{18}^{\psi}]^3$,   as $N$ is even we must have the equality
\be A_{\rm IR}=A_{\rm UV} +18 \, m +  9^3 \, n = A_{\rm UV} +  9 \,k\ , \qquad   m,n,k \in {\mathbbm Z}\ , 
\ee
i.e., an equality {\it modulo} $9$,
if   $ {\mathbbm Z}_{18}^{\psi}$ is to remain unbroken in the infrared.    But
\be     A_{\rm UV} =2 \cdot 20 = 4 \mod 9\ .
\ee
The consideration of  ${\mathbbm Z}_{18}^{\psi}\,[SU(2)]^2$ and   $[ {\mathbbm Z}_{18}^{\psi}]^3$   anomaly matching  is consistent with the assumption of the bi-fermion condensate (\ref{SU(3)cond})   (the reduction   $ {\mathbbm Z}_{18} \longrightarrow {\mathbbm Z}_{2}$)
and with consequent nine-fold degeneracy of the vacua.    The 1-form center symmetry gauging and the mixed anomaly alone, give instead a  weaker condition (\ref{weaker}).

\subsubsection{$N_f = 4$}

The symmetry of the  $N_f = 4$ model  is 
\be   G_f=    SU(4)\times {\mathbbm Z}_{24}^{\psi} \ .
\ee
The 1-form gauging of ${\mathbbm Z}_2^{C}$  breaks the discrete chiral symmetry as
\be    {\mathbbm Z}_{24}^{\psi} \longrightarrow    {\mathbbm Z}_{8}^{\psi} \ .
\ee
In this case,   the bi-fermion condensate (\ref{SU(2)cond}) of the form,
\be     \langle  \psi^{[A} \psi^{B]}  \rangle   \ne 0  \ , \qquad  A,B=1,2    \label{SU(2)condBis}
\ee
or 
\be     \langle  \psi^{[A} \psi^{B]}  \rangle   \ne 0  \ , \quad  (A,B=1,2) \ , \qquad   \langle  \psi^{[C} \psi^{D]}  \rangle   \ne 0  \ , \quad  (C,D=3,4)\ ,    \label{SU(2)condTris}
\ee
if it occurs,   break  the symmetry as
\be       SU(4)\times {\mathbbm Z}_{24}^{\psi}    \longrightarrow    SU(2) \times SU(2) \times {\mathbbm Z}_{2}^{\psi}   \ . 
\ee
There are $15-6=9$  NG bosons. Again no massless  baryons are required. 

Thus for $N_f=4$, the discrete $ {\mathbbm Z}_{24}^{\psi}$ would be broken  to  $ {\mathbbm Z}_{8}^{\psi}$  due to the 1-form center gauging, whereas it would be broken more strongly to 
$ {\mathbbm Z}_{2}^{\psi}$,  if the condensate (\ref{SU(2)condBis}) or   (\ref{SU(2)condTris})  is to form. 

As for the conventional discrete anomaly is concerned,  it has an anomaly at UV:
 \be  A_{\rm UV} \big({\mathbbm Z}_{24}^{\psi}\,[SU(2)]^2 \big)=20   =-4 \mod 24 \ , \ee
 implying a discrete symmetry breaking to ${\mathbbm Z}_{4}^{\psi}$ and a six-fold vacuum degeneracy, at least.  This does not exclude that it is broken 
 more strongly, as expected from the bi-fermion condensate formation.
 
 As for the $[{\mathbbm Z}_{24}^{\psi}]^3$,  there is an UV anomaly
 \be A_{\rm UV} \big( [{\mathbbm Z}_{24}^{\psi}]^3\big) = 4 \cdot 20\ . \ee
This must match to the IR modulo   $\gcd{(24,\frac{24^3}{8})}=24$, a requirement satisfied by the  reduction of the symmetry to ${\mathbbm Z}_{4}^{\psi}$.

\subsubsection{$5  \le N_f \le 10$}

For larger  $N_f$, the symmetry of the system  is 
\be    G_f=    SU(N_f)\times   {\mathbbm Z}_{6 N_f}^{\psi} \ .
\ee
For illustration let consider the case  $N_f=5$.  The discrete 
 ${\mathbbm Z}_{30}^{\psi} $ symmetry is broken by the 1-form gauging to 
 ${\mathbbm Z}_{10}^{\psi} $,  implying some condensates to occur in the infrared.    
A condensate of the form, (\ref{SU(3)cond}), would break the global symmetry as, 
\be    SU(5)\times   {\mathbbm Z}_{30}^{\psi}    \longrightarrow     SU(2) \times  SU(3)\times   {\mathbbm Z}_{2}^{\psi}  \ .
\ee
It would be a hard problem to find a set of massless baryons saturating the anomaly triangles  associated with this low-energy symmetries. 

It is possible, however, that the system instead chooses to produce condensates of the form, 
\be     \langle  \psi^{[A} \psi^{B]}  \rangle \ne 0  \ , \quad   A,B=1,2\ ;  \qquad   \langle  \psi^{[C} \psi^{D]}  \rangle  \ne 0 \ , \quad   C,D=3,4\ .  \label{consistent}
\ee
In this case the symmetry breaking pattern is 
\be      SU(5) \times   {\mathbbm Z}_{30}^{\psi}   \longrightarrow   SU(2) \times  SU(2) \times   {\mathbbm Z}_{2}^{\psi}  \ ,
\ee
and the low-energy theory is described by 
$ 24-6=18 $
massless NG bosons.  No massless baryons are required, and are expected to appear. 

Finally, let us check the conventional anomalies involving $ {\mathbbm Z}_{30}^{\psi}$.  In the UV, the anomalies are
\bea    A_{\rm UV}  \big({\mathbbm Z}_{30}^{\psi}  \,[{\rm grav}]^2\big) &=&   5 \cdot 20 =100\neq 0 \mod 30\ ,\nonumber \\
    A_{\rm UV} \big([{\mathbbm Z}_{30}^{\psi}]^3\big) &=& 5 \cdot 20 =100 \neq 0 \mod 30\ ,
\eea
and would not match in the UV and IR, implying a (at least) partial breaking of $ {\mathbbm Z}_{30}^{\psi}$.  However, if  the discrete symmetry is reduced to ${\mathbbm Z}_{2}^{\psi}$ by the bi-fermion condensates, then 
the UV   ${\mathbbm Z}_{2}^{\psi}$  anomalies vanish
\bea  
 A_{\rm UV} \big({\mathbbm Z}_{2}^{\psi}   \,[{\rm grav}]^2\big) &=&   5 \cdot 20 =100 = 0 \mod 2\ ,\nonumber \\
 A_{\rm UV} \big([{\mathbbm Z}_{2}^{\psi}]^3\big) &=& 5 \cdot 20 =100 = 0     \mod 2 \ ,
\eea
and no contradiction arises. 

\subsection{$SU(N)$ generalizations \label{sec:special}}

We now consider the  general case of  $SU(N)$ ($N$ even) theory with $N_f$  left-handed fermions $\psi$  in the self-adjoint, totally antisymmetric, one-column (of height $n=\frac N2$)  representation.
The first coefficient of the beta function is
\be  b_0=   \frac{ 11 N -  2 N_f T_R }{3}\ .
\ee
For simplicity, we shall limit ourselves to the single flavor ($N_f =1$) case below.  The generalization to general $N_f $ is quite straightforward.
For $SU(N)$   one finds that  the twice Dynkin index  (see Appendix~\ref{Dynkin}) is given by 
\be    2 \, T_R =  { {  N-2}    \choose {{(N-2)/2}}} \ :\ee
$2 T_R$ and $d(R)$ are given for some even values of $N$ in     Table~\ref{dynkins}.
\be \begin{tabular}{|c |c cccc| }
\hline
$N$ & $4$ & $6$ & $8$ & $10$ & $12$     \\
 \hline
$2 \, T_R$ & $2$  & $6$ & $20$ & $70$ & $252$   \\
$d(R)$  &  $6$ &  $20$ &  $70$ &  $252$ & $924$  \\  
   \hline
\end{tabular}   \ .
\label{dynkins}
\ee

Thus  $SU(4)$, $SU(6)$, $SU(8)$,  $SU(10)$  models  with $N_f=1$  are asymptotically free, $SU(12)$ and higher are not.  We shall limit  ourselves to some of the asymptotically free theories.

$\psi$ is neutral with respect to the  ${\mathbbm Z}_{\frac N2}^{C}$ symmetry, therefore the system possesses an exact 
1-form: 
\be     
{\mathbbm Z}_{\frac N2}^{C} \,:\quad e^{ i \oint A } \to      e^{\frac{  2\pi i  }{ N}  k }   \,   e^{ i \oint A }   \ ,  \qquad        k=2,4,\ldots  N \ .
\ee
  At the same time,  the anomaly-free global  discrete symmetry 
is: 
\be {\mathbbm Z}_{2 T_R}\,:\quad   \psi \to  e^{\frac{2\pi i}{2 T_R}j}\, \psi \ ,  \qquad        j=1,2,\ldots 2 T_R\ .
\ee
We are interested to find out how this discrete symmetry is realized in the infrared, and 
what the 1-form gauging of the center symmetry has to tell about it.  

We introduce a 1-form gauge fields $(B_c^{(2)}, B_c^{(1)})$  such that 
\be          \frac{N}{2}   B_c^{(2)} =d B_c^{(1)}\ .
\ee
The anomaly can be evaluated from the Stora-Zumino  $6$D Abelian anomaly \cite{StoraZumino}  
\bea   && {1
 \over 24  \pi^2}{\tr}_{ R
}\big({\tilde F}-B^{(2)}_c   -   d A_{\psi}  \big)^3 \nonumber \\
&&  =    {  2 T_R  \over 8\pi^2}    {\tr} \big( {\tilde F}-B^{(2)}_c\big)^2 \wedge dA_{\psi} + \ldots   \nonumber \\
 &&  =    {  2 T_R  \over 8\pi^2}   {\tr}  {\tilde F}^2  \wedge dA_{\psi}  -  { 6 N \over 8\pi^2}   (B_c^{(2)})^2  \wedge dA_{\psi}   + \ldots   \label{formula}
\eea
so that the $5D$ effective action reads, 
\be     S^{5D}   =   {  2 T_R  \over 8\pi^2}   {\tr}  {\tilde F}^2  \wedge  A_{\psi}  -  { 6 N \over 8\pi^2}   (B_c^{(2)})^2  \wedge  A_{\psi}   + \ldots   \label{formula5D}
\ee  
The first term is clearly trivial, as
\be   {1 \over 8\pi^2}   \int  { \tr } {\tilde F}^2  \in   {\mathbbm Z}, \qquad   A_\psi  = d A_\psi^{(0)}\;, \ee 
and 
\be   \delta A_\psi^{(0)} =   \frac{ 2\pi k}{  2 T_R }\ ,\qquad k=1,2,\ldots, 2T_R\ .
\ee
which corresponds to the 
${\mathbbm Z}_{2 T_R}$ transformation of the $\psi$ field in four dimensional action.  As for the second term of (\ref{formula5D}),  as
\be    -   { 2 T_R   \, N    \over 8\pi^2}   \int    (B_c^{(2)})^2  \in   -    2 T_R   N   \Big(\frac{2}{N}\Big)^2  {\mathbbm Z} =   -  2 T_R    \frac{4}{N} \,  {\mathbbm Z} \   \label{Factor}
\ee
the phase of the partition function is transformed by 
\be    - \frac{2\pi k  }{2 T_R}     2 T_R   \frac{4}{N} {\mathbbm Z}   =       - {2 \pi  k}  \frac{4}{N} {\mathbbm Z}\ , \qquad k=1,2,\dots, 2T_R\ ,
\ee
under ${\mathbbm Z}_{2 T_R}$: one sees that 
the 1-form gauging of the center ${\mathbbm Z}_{N/2}^{C}$ has the effect of making ${\mathbbm Z}_{2 T_R}$ anomalous, in  general.  Stated differently, there is a mixed 
 't Hooft anomaly between the 1-form ${\mathbbm Z}_{\frac N2}^{C}$ gauging and 0-form ${\mathbbm Z}_{2 T_R}$ symmetry.  Its consequence depends on $N$
 in a nontrivial fashion:
\begin{description}
\item[(i)]
For $N=4$,   the mixed anomaly disappears, as 
\be   \frac{4}{N}=1\ . 
\ee
 \item[(ii) ]  For $N=4 \ell$, $\ell \ge 2$, 
\be        \frac{4}{N}=      \frac{1}{ \ell}\ , 
\ee
therefore  the discrete  symmetry is  broken to  
\be    {\mathbbm Z}_{2T_R}^{\psi} \longrightarrow  {\mathbbm Z}_{\frac{2T_R}{\ell}}^{\psi}  \; 
\ee   
generated by 
\be     \psi \to    e^{\frac{2\pi i }{ 2 T_R}  k}\psi  \ , \qquad   k= \ell,  2\ell, 3\ell, \ldots,  2 T_R\; .
\ee
Note that for  $N=4\ell$,     $2 T_R$  is an integer multiple of $\ell $ (see  Appendix~\ref{Dynkin}).
  \item[(iii)] 
  For  $N=4\ell +2$,  
\be  2 T_R   \cdot  \frac{4}{N}=     2 T_R   \cdot  \frac{2}{2 \ell+1}\ , 
\ee
therefore  the breaking of the discrete  symmetry is 
\be    {\mathbbm Z}_{2T_R}^{\psi} \longrightarrow   {\mathbbm Z}_{\frac{2  T_R}{2\ell+1}}^{\psi}  \ ;
\ee
only the transformations 
\be A_\psi =   d  A_\psi^{(0)}\;,    \quad   \delta A_\psi^{(0)}=    \frac{2\pi k}{ 2 T_R} \ ,\quad k=  2 \ell+1, 4\ell+2, \ldots, 2 T_R\;  \ee
remain invariant.
Note that for $N$ of the form, $N=4\ell+2$,      $2\ell+1 $ is a divisor of  $2 T_R$  (see Appendix~\ref{Dynkin}).

\end{description}

  Concretely,  for $SU(4)$ there are no mixed anomalies.  $SU(6)$ case has been studied in detail above:
  the discrete symmetry  (for $N_f=1$)  is broken as (\ref{reduced})
 by 1-form gauging.
 For $SU(8)$  the effect of the 1-form center symmetry gauging is
  \be   {\mathbbm Z}_{20}^{\psi} \longrightarrow   {\mathbbm Z}_{10}^{\psi} \ ,
 \ee
 and for $SU(10)$ is
 \be   {\mathbbm Z}_{70}^{\psi}\longrightarrow  {\mathbbm Z}_{14}^{\psi} \ .
 \ee

The Fermi statistics  allows, for $N$ multiple of $4$, (e.g.,   $N=4$ or $N=8$),  a bi-fermion condensate
\be    \brc \psi \psi \ckt\ ,   
\ee
which is gauge invariant.  If such a condensate indeed forms the discrete symmetry is broken more strongly, as 
 \be   {\mathbbm Z}_{20}^{\psi} \longrightarrow   {\mathbbm Z}_{2}^{\psi} \ ,\label{bfcond}
 \ee
 
For $N$ of the form, $N= 4\ell +2$, $\ell \in {\mathbbm Z}$, a bi-fermion Lorentz invariant condensate cannot be gauge invariant. 
As discussed in $SU(6)$ case,  it is possible that in these cases dynamical Higgs phenomenon occurs, with gauge noninvariant 
bi-fermion condensate in the adjoint representation of $SU(N)$. The system can dynamically Abelianize.   The discrete symmetry is again
broken to ${\mathbbm Z}_{2}^{\psi}$.

Finally, let us check the conventional anomalies involving ${\mathbbm Z}_{2T_R}^{\psi} $.  
In the UV, the anomalies are
\bea    A_{\rm UV}  \big({\mathbbm Z}_{2T_R}^{\psi}  \,[{\rm grav}]^2\big)  &=&   1 \cdot   d(R)  \ne  0  \mod  2T_R \ ,\nonumber \\
    A_{\rm UV} \big([{\mathbbm Z}_{2T_R}^{\psi}]^3\big) &=& 1 \cdot d(R)     \ne  0  \mod  2T_R   \,.
\eea
except for $N=4$  (see Table~\ref{dynkins}).     Thus the conventional discrete anomaly matching requirement implies that 
some condensate forms in the infrared, breaking ${\mathbbm Z}_{2T_R}^{\psi} $  spontaneously.  The assumption of bi-fermion condensate and consequent spontaneous breaking of  ${\mathbbm Z}_{2T_R}^{\psi} $, (\ref{bfcond}), is compatible with the discrete anomaly matching condition. 

~~~

Let us discuss these results from the point of view of the fractional instantons. 
If no matter fields are present in the system, one can compactify the ${\mathbb R}^4$ space on a 4-torus ${\mathbb T}^4$ and insert one unit of  't Hooft flux in the first 2-torus $(x_1,x_2)$ and another unit  in the second 2-torus $(x_3,x_4)$. This object \cite{tHooft:1979rtg}-\cite{vanBaal:1982ag}, sometimes called ``toron'', has topological charge equal to $\frac 1N$ that of an ordinary instanton. In general we can insert $n_{12}$ units of  't Hooft flux in the first 2-torus  and $n_{34}$ units in the second and this object has topological charge $\frac{n_{12} n_{34} }{N}$. If the instanton breaks a certain chiral  symmetry to a discrete subgroup $U(1) \longrightarrow {\mathbbm Z}_{M}$ this is due to the presence of $M$ fermion zero modes in the instanton background. A toron  has a  smaller amount of zero modes, precisely $\frac MN$ due to the index theorem and  thus the discrete symmetry is  broken further to ${\mathbbm Z}_{\frac MN}$.
It is known \cite{Komargodski:2017smk,Popp} that gauging the 1-form center symmetry is equivalent to putting the theory on a nontrivial background with fractional instanton number.

Here we are interested in theories with matter fields, but with some residual center symmetry. This means that fractional instantons can be constructed, but not of the one of  the minimal charge. 
For the   $SU(N)$ ($N$ even) theory with $N_f$  left-handed fermions $\psi$  in the self-adjoint representation the remaining center symmetry is  ${\mathbbm Z}_{\frac N2}^C$ which means that only {\it even} numbers of fluxes are allowed on each $2$-torus. With $n_{12}=n_{34}=2$ units of fluxes we  have a toron with charge $\frac{4}{N}$. This can be combined with any integer number of instanton charge to construct the minimal possible instanton charge  
$\frac{\gcd{(4,N)}}{N}$ and this is $\frac{4}{N}= \frac{1}{\ell}$ for $N = 4 \ell $ and $\frac{2}{N}= \frac{1}{2\ell+1}$ when $N = 4 \ell + 2$.
The symmetry is then  broken as
\beq
U(1)_{\psi} \longrightarrow {\mathbbm Z}_{2 T_R N_f }^{\psi} \longrightarrow {\mathbbm Z}_{2 T_R N_f \,  \frac{\gcd{(4,N)}}{N}}^{\psi}
\eeq
first by the ABJ anomaly and instantons,  and then by the gauging of the 1-form symmetry. This result agrees with what was found above by use of the  ${\mathbbm Z}_{\frac N2}^C$  gauge fields $(B_c^{(2)}, B_c^{(1)})$. More about these issues at the end, see Discussion (Section~\ref{conc}).

\section{Adjoint QCD}
\label{aqcd}

$SU(N)$  theories with $N_f$ Weyl fermions  $\lambda$  in the adjoint representation have been the object of intense study, 
 and our comments here will be brief.   In this model, the color center 1-form ${\mathbbm Z}_N^{C}$ symmetry is exact, therefore can be entirely gauged.
The system possesses also a nonanomalous $0$-form discrete chiral symmetry, 
\be    {\mathbbm Z}_{2 N_f N}^{\lambda}: \quad   \lambda \to  e^{\frac{2\pi i  }{2 N_{ f} N} k} \lambda\ , \qquad  k=1,2,\ldots,  2 N_f N\ .\label{0formdisc}
\ee
We introduce a set of gauge fields
\begin{itemize}
\item $A_{\lambda}$: \,\, $ {\mathbbm Z}_{2 N_f N}^{\lambda}$ \,\, 1-form gauge field,
      to formally describe  (\ref{0formdisc});
\item $B^{(2)}_c$:    \,\,$\mathbb{Z}_{N}^{C}$ \,\, 2-form gauge field.
\end{itemize}
The Abelian $6$D anomaly is
\bea   &&    \frac{1}{24 \pi^2}     \int       {\tr}_{\rm adj}  \big(  {\tilde F} -  B_c^{(2)} -   d   A_{\lambda} \big)^3   \nonumber \\
&&=       \frac{2 N  N_f   }{8\pi^2}     \int    {\tr}_{}  (  {\tilde F} -  B_c^{(2)} )^2  \wedge d A_{\lambda} + \ldots   \nonumber \\
&&
=     \frac{2 N  N_f   }{8\pi^2}      \int   {\tr}_{} {\tilde F}^2 \wedge d A_{\lambda}  -  \frac{2 N^2  N_f   }{8\pi^2}     \int   (B_c^{(2)} )^2   \wedge d A_{\lambda} \ldots  \ .
\eea
As
\be  A_{\lambda} =  d  A_{\lambda}^{(0)}\;, \qquad    \delta A_{\lambda}^{(0)}  \in     \frac{2 \pi i}{ 2 N N_f}  {\mathbbm Z}
\ee
the first term is trivial  (conserves  $ {\mathbbm Z}_{2 N N_f}^{\lambda}$);   the second term gives 
\be \Delta S ( \delta A_{\lambda}^{(0)} )    \in     \frac{2 \pi i}{N} {\mathbbm Z}\ ,
\ee   
breaking the chiral discrete symmetry as 
\be    {\mathbbm Z}_{2 N N_f}^{\lambda} \longrightarrow      {\mathbbm Z}_{2 N_f}^{\lambda}  \label{adjqcd}
\ee
in agreement with \cite{GKKS,ShiYon,AnbPop1}.  In this case the matter fields have no charge under the center of the gauge group and so  the torons can have the minimal topological charge $\frac 1N$ of that of the instanton, hence the breaking (\ref{adjqcd}).

 Let us briefly discuss the case of $SU(2)$, $N_f=2$ theory.  The discrete chiral symmetry ${\mathbbm Z}_{8}^{\lambda}$ is in this case broken by the 1-form  ${\mathbbm Z}_{2}^{C}$ gauging to 
 as 
\be
 {\mathbbm Z}_{8}^{\lambda}\longrightarrow   {\mathbbm Z}_{4}^{\lambda} \ , \label{1formgauging}
\ee
with 
\be
 {\mathbbm Z}_{4}^{\lambda}:\quad  \lambda \to e^{\frac{2\pi i  }{8}  k}\,  \lambda \ , \qquad  k=2,4,6, 8 \ .  
\ee
In particular it means that the discrete chiral transformations
\be  \lambda \to  e^{\pm \frac{ 2\pi i}{8}} \lambda\ ,  \label{mixedBis}
\ee
which is an invariance of the standard $SU(2)$ theory,  becomes anomalous under the gauging of  ${\mathbbm Z}_{2}^{C}$, i.e., 
in the quantum  $SU(2)/{\mathbbm Z}_{2}$ theory.

A familiar assumption about the infrared dynamics of this system  is  \cite{Shifman}    that a condensate
\be  \langle \lambda^{\{I} \lambda^{J\}} \rangle \ne 0 \ , \qquad    SU_f(2)  \longrightarrow SO_f(2)\;   \label{familiar}  \ee
($I,J=1,2$ being the flavor $SU_f(2)$ indices)  forms in the infrared.  
 It would break the discrete chiral symmetry as   $ {\mathbbm Z}_{8}^{\lambda} \to {\mathbbm Z}_{2}^{\lambda} $,  which leaves four-fold degenerate vacua.  
Note that this symmetry breaking is stronger than that would follow from the 1-form gauging, (\ref{1formgauging}).  The fact that the vacuum breaks the symmetry further with respect to what is expected from  (\ref{1formgauging}) is a common feature, seen also in the previous Section \ref{sa}, and is perfectly acceptable.

Anber and Poppitz (AP) \cite{AnbPop1} however propose  that the system instead develops a four-fermion condensate of the form
\be    \langle  \lambda \lambda \lambda \lambda  \rangle \ne 0\ , \quad {\rm with } \quad  \langle \lambda \lambda \rangle  =0\ , \label{AnbPop}
\ee
in the infrared,   which breaks  $ {\mathbbm Z}_{8}^{\lambda} $ spontaneously to  ${\mathbbm Z}_{4}^{\lambda}$,  
leaving only doubly degenerate vacua
 with unbroken   $SU_f(2)$ symmetry.  
 They assume that  massless baryons of spin $\tfrac{1}{2}$ 
 \be    B \sim   \lambda \lambda \lambda \;\label{AnbPop2}
 \ee
 which is necessarily a doublet of unbroken $SU_f(2)$, appear in the infrared spectrum. 
   It is shown that all the conventional 't Hooft and Witten anomaly matching conditions are met by
 such a vacuum and with such low-energy degrees of freedom. The action of the broken $  {\mathbbm Z}_{8}^{\lambda}/ {\mathbbm Z}_{4}^{\lambda}$ is seen to be realized in the infrared as
 a transformation between the two degenerate vacua,
 \be     \langle  \lambda \lambda \lambda \lambda  \rangle \to -  \langle  \lambda \lambda \lambda \lambda  \rangle\ .      \label{AnbPopZ2}
 \ee
 Most significantly Anber and Poppitz note  that the above hypothesis is consistent with what is expected from the gauging of the $1$-form ${\mathbbm Z}_2^{C}$ center symmetry,  (\ref{1formgauging}), this time just with the minimal amount of breaking necessary.

Let us check  the conventional anomalies associated with the discrete symmetry in the Anber-Poppitz scenario.  Assuming the  condensates of the form (\ref{AnbPop}), the anomalies associated with the unbroken  ${\mathbbm Z}_{4}^{\lambda}$ symmetries must be considered.  The anomalies  $ {\mathbbm Z}_{4}^{\lambda}\,[ SU_f(2)]^2 $  and  
 ${\mathbbm Z}_{4}^{\lambda}\,[{\rm grav}]^2 $  have been already verified  in  \cite{AnbPop1}  to match  in the UV and in the IR,  therefore 
 only the   $[{\mathbbm Z}_{4}^{\lambda}]^3$ anomaly
 remains to be checked. In the UV  $\lambda$  contributes
 \be  A_{\rm UV}\big([{\mathbbm Z}_{4}^{\lambda}]^3\big)  = N_f \cdot d({\rm adj}) = 2 \cdot 3 =2 \mod 4\ , \ee
 whereas in the  IR    the baryons $B$ gives 
  \be  A_{\rm IR}\big([{\mathbbm Z}_{4}^{\lambda}]^3\big) = N_f  \cdot 3^3 = 2 \cdot 27    =2 \mod 4\ ,   \ee 
 therefore the matching works.

As in any anomaly-matching argument, these considerations only tell that a particular dynamical scenario (in this case,  (\ref{AnbPop})-(\ref{AnbPopZ2}))
is consistent, but not that such a vacuum is necessarily realized. 
It would be important to establish which between  the familiar  $SO_f(2)$ symmetric vacuum and  the proposed   $SU_f(2)$ symmetric one,  is actually realized in the infrared, 
 e.g.,  by using the lattice simulations.

The adjoint QCD has been discussed extensively in the literature,  by compactifying one space direction to $S^1$ and by using controlled semi-classical analysis \cite{Unsal2007},  by direct lattice simulations \cite{Debbio}, and more recently, by applying the 1-form center symmetry gauging and using mixed anomalies \cite{GKKS,ShiYon,AnbPop1}.  For more general approach, see \cite{Shifman}, and for more recent work on adjoint QCD, see 
\cite{Popp,WanWang}.

 For $N_f=1$ the system reduces to ${\cal N}=1$ supersymmetric Yang-Mills theory, where 
a great number of nonperturbative results are available \cite{AffleckDineSei,NSV,AmatiKMV,DaviesHollo}.  Note that for $N_f=1$  $SU(N)$ theory,  the breaking of the discrete symmetry (\ref{adjqcd}) due to the 1-form gauging implies an $N$ fold vacuum degeneracy,  in agreement with the well-known result,  i.e.,  the Witten index of pure ${\cal N}=1$ $SU(N)$ Yang-Mills.  

Another possibility is to start from the ${\cal N}=2$ supersymmetric $SU(2)$ Yang-Mills theory,
where many exact results for the infrared effective theory are  known \cite{SW,Tachikawa}. It can be deformed to ${\cal N}=1$ 
theory
by a mass perturbation, yielding a confining, chiral symmetry breaking  vacua.  For the exact calculation of gauge fermion condensates $\brc \lambda \lambda\ckt$  from this viewpoint, see \cite{FinnelPouliot,RiccoK}. 
The pure ${\cal N}=2$ theory can also  be deformed directly to ${\cal N}=0$ \cite{Cordova}, to give indications about $N_f=2$ adjoint QCD.

\section{QCD with quarks in a two-index representation
\label{two}}

Consider now  $SU(N)$, $N$ even,  with  $N_f$ pairs of  "quarks" in symmetric (or antisymmetric)  representations.  Namely the left-handed matter fermions are either 
\be     \psi, {\tilde \psi }  =  \yng(2) \oplus   {\bar {\yng(2)} }    \label{symmquarks}
\ee
or 
\be 
   \psi, {\tilde \psi }  =   \yng(1,1) \oplus   {\bar {\yng(1,1)} }  \label{antisymmquarks}
\ee
(the quarks in standard QCD are  in    $  \yng(1) \oplus   {\bar {\yng(1)} }  $).
The first beta function coefficients are
\be  b_0= \frac{11 N -   2 N_f  (N \pm 2)}{3} \ .
\ee
The  $k= \frac{N}{2}$ element of the  center ${\mathbbm Z}_N$ does not act on $\psi$'s, i.e.,  there is an exact 
\be  {\mathbbm Z}_2^{C}  \subset {\mathbbm Z}_N^{C}
\ee
center symmetry.\footnote{This aspect has been considered by Cohen \cite{Cohen}, in particular in relation with the possible existence of an order parameter for confinement.}
On the other hand, there is a discrete  axial symmetry 
\be {\mathbbm Z}_{2 N_f (N \pm 2)}^{\psi}\,:\quad    \psi   \to    e^{\tfrac{2\pi i  }{ 2 N_f (N \pm 2)} } \, \psi\ ,\quad   {\tilde \psi }   \to    e^{\tfrac{2\pi i  }{ 2 N_f (N \pm 2)} } \,{\tilde \psi } \,,
\ee
preserved by instantons.  The $\pm$ signs above refer to two cases  Eq.~(\ref{symmquarks}) and Eq.~(\ref{antisymmquarks}), respectively.

Let us consider for simplicity $N_f=1$ and consider a 1-form gauging of the exact ${\mathbbm Z}_2^{C}$ .   The external background fields are 
\begin{itemize}
\item $A_{\psi}$: $ {\mathbbm Z}_{2 (N \pm 2)}^{\psi}$ 1-form gauge field,
\item $B^{(2)}_c$: $\mathbb{Z}_{2}^{C}$ 2-form gauge field.   
\end{itemize}
The last satisfies 
\be    2 B_c^{(2)} = d B_c^{(1)}\ , \qquad   B_c^{(1)}  \to  B_c^{(1)} + 2 \lambda\ ,  \qquad  B_c^{(2)} \to B_c^{(2)}  + d \lambda   
\ee
The $6$D anomaly is
\bea   &&    \frac{1}{24 \pi^2}         {\tr}_{R_{\psi}}  \big(  {\tilde F} -  B_c^{(2)}      -    d   A_{\psi} \big)^3   +       \frac{1}{24 \pi^2}         {\tr}_{R_{\tilde{\psi}}}  \big(  {\tilde F} -  B_c^{(2)}        +    d   A_{\psi} \big)^3  + \dots \nonumber \\
&&=     -   \frac{2(N \pm 2)}{8\pi^2}         {\tr}_{}  (  {\tilde F} -  B_c^{(2)} )^2  \wedge  d A_{\psi} + \dots
\nonumber \\
&&
=   -   \frac{2(N \pm 2)}{8\pi^2}       {\tr} {\tilde F}^2 \wedge d A_{\psi}    +   \frac{2N (N \pm 2)}{8\pi^2}      (B_c^{(2)} )^2     \wedge d A_{\psi}    + \dots  \label{secondterm}  \eea
Now
\be   \frac{2(N \pm 2)}{8\pi^2}       \int   {\tr}_{} {\tilde F}^2 \in  2 (N \pm 2)    {\mathbbm Z}  \ ,  \ee
\be     A_{\psi} =d A_{\psi}^{(0)}\;, \qquad           \delta A_{\psi}^{(0)}  \in \frac{2\pi }{2(N \pm 2)}   {\mathbbm Z}_{2(N \pm 2)}  \ ,
\ee
so the first term is trivial.   By using 
\be     \frac{1}{8\pi^2}  \int     (B_c^{(2)} )^2 =   \frac{1}{4} {\mathbbm Z}  \ ,
\ee
the second term gives an anomaly
\be   A  =  2\pi  \frac{N}{4}  \,   {\mathbbm Z}  \ ,
\ee
This means that for $N= 4\ell$ there is no anomaly, whereas  for  $N= 4\ell + 2$ the 1-form gauging breaks the discrete symmetry as 
\be
   {\mathbbm Z}_{2 (N \pm 2)}^{\psi} \longrightarrow    {\mathbbm Z}_{N \pm 2}^{\psi}
\label{bsr}
\ee
with the  subgroup 
\be {\mathbbm Z}_{N \pm 2}^{\psi}\,:\quad   \psi   \to    e^{\tfrac{2\pi i   }{ 2(N \pm 2)} \ell  } \, \psi\ ,\quad   {\tilde \psi }   \to   e^{\tfrac{2\pi i   }{ 2(N \pm 2)}  \ell }  \,{\tilde \psi } \ ,
\qquad   \ell = 2,4,\ldots  2(N \pm 2)
\ee
that remains nonanomalous.

We can construct a toron with $n_{12}=n_{34}=\frac N2$ units of fluxes and thus topological charge $\frac{N}{4}$. For $N$ multiple of $4$ this is not fractional and thus we have no mixed anomaly; for $N = 4\ell + 2$, it can be combined with a suitable number of instantons to obtain the minimal possible fractional charge $\frac 12$ and this explains the breaking (\ref{bsr}).

These results are consistent with the assumption that in the IR   the condensate
\be  \langle  \psi  {\tilde \psi} \rangle  \ne 0     \label{itself}
\ee
forms, even though the bi-fermion condensate itself (\ref{itself}) breaks the discrete symmetry more strongly,
\be     {\mathbbm Z}_{2 (N \pm 2)}^{\psi}\longrightarrow   {\mathbbm Z}_{2}^{\psi}\ .
\ee

\section{Chiral models with $ \tfrac{N-4}{k}  $   $\psi^{\{ij\}}$'s and    $\tfrac{N+4}{k}$   ${\bar  \chi_{[ij]} }$'s    }
\label{chiral}

Let us consider now $SU(N)$ gauge theories with Weyl fermions in the complex representation,  $ \tfrac{N-4}{k}  $   $\psi^{\{ij\}}$'s and    $\tfrac{N+4}{k}$   ${\bar  \chi_{[ij]} }$,
 \be   \frac{N-4}{k}  \,\,   \yng(2)   \oplus   \frac{N+4}{k}     \,\,    {\bar  {\yng(1,1)} }  \ ,   \ee
where $k$ is a common divisor of $(N-4, N+4)$ and  $N\ge 5$.  With this matter content the gauge anomaly  cancels.
Asymptotic freedom requirement 
\be     11 N -   \frac{2}{k} (N^2-8) >0\ , 
\ee
leaves a plenty of possibilities for $(N, k)$.   Two particularly simple models which we  analyze  in the following are:
\begin{description}
  \item[(i)] $(N, k) = (6,2)$: $SU(6)$ theory with
  \be   \yng(2) \oplus    5\,\, {\bar {\yng(1,1)}}\ ;\label{model1}
  \ee 
  \item[(ii)]   $(N, k) = (8,4)$: $SU(8)$  model with  
  \be      \yng(2) \oplus  3\,\, {\bar {\yng(1,1)}}\ .\label{model2}
  \ee
\end{description}

\subsection{$SU(6)$ theory with  \,\underline{$21$}\,$\oplus$\,$5$\,$ \times $\,\underline{${15}$}$^{*}$}

Classical continuous flavor symmetry group  is 
\be SU(5) \times  U(1)_{\psi}\times U(1)_{\chi}\ . \ee 
The chiral anomalies are:
\bea U(1)_{\psi} \,[SU(6)]^2 &=&\frac{T_{\tiny \yng(2)}}{T_{\tiny \yng(1)}}=N+2=8 \ , \nonumber \\
 U(1)_{\chi} \,[SU(6)]^2&=&\frac{5 T_{\tiny \bar{\yng(1,1)}}}{T_{\tiny \yng(1)}}=5(N-2)=20 \ ,
\eea
meaning that the charges with respect to the unbroken $U(1)_{\psi\chi}\subset  U(1)_{\psi}\times U(1)_{\chi} $  symmetry are
\be  (Q_{\psi}, Q_{\chi}) = (5,- 2)\ .\ee
The system has unbroken discrete groups also:
\be U(1)_{\psi} \longrightarrow {\mathbb Z}_8^{\psi}\ , \qquad U(1)_{\chi}\longrightarrow {\mathbb Z}_{20}^{\chi} \ .\ee
One might wonder if   a subgroup of 
$ \Z_8^{\psi} \times\Z_{20}^{\chi} $  is contained in $U(1)_{\psi\chi}$.  In fact,  $U(1)_{\psi\chi}$ transformations
\be  \psi \to e^{  5   i \alpha} \psi\ , \qquad \chi \to e^{- 2 i  \alpha} \chi\ , 
\ee
with
\be      \alpha=   \frac{2\pi  k}{40}\ , \qquad k=1,2,\ldots, 40\ ,
\ee
generate  the subgroup 
\be      \psi \to  e^{ \frac{ 2\pi   i}{  8 } k  } \psi\ ,\qquad  \chi \to  e^{  -   \frac{2\pi   i} {20}  k } \chi\;\label{Z40}
\ee
of   ${\mathbbm Z}_8^{\psi} \times {\mathbbm Z}_{20}^{\chi} $.
The anomaly-free symmetry subgroup of $ U(1)_{\psi}\times U(1)_{\chi} $  is 
\be  \frac{U(1)_{\psi\chi}\times {\mathbbm Z}_8^{\psi} \times {\mathbbm Z}_{20}^{\chi} }{{\mathbbm Z}_{40}} \sim  U(1)_{\psi\chi}\times \Z_4\;.
 \label{global} \ee 
(see (\ref{quotient}) and (\ref{Z40Bis}) below).
Actually, by considering the overlap with color center  and $SU_f(5)$ center,  the correct anomaly-free  symmetry  group is:\footnote{Here we do not gauge the full denominator of the global group, but only the exact subgroup of the center symmetry. This will be done, for a simpler chiral gauge theory, in \cite{BKT}.}
\be 
 \frac{SU(5) \times  U(1)_{\psi}\times U(1)_{\chi}}{\Z_6^{C} \times\Z_5^f} \longrightarrow \frac{SU(5) \times  U(1)_{\psi\chi}\times \Z_4}{\Z_6^{C} \times\Z_5^f} \ . \label{full}
 \ee

Let us first check the  't Hooft  anomaly matching condition with respect to the continuous global symmetries, assuming that the vacuum possesses the full symmetry, (\ref{full}).
The anomaly coefficients in the UV  are 
\[  
  \begin{tabular}{|c|c|c|c|c|c|}
\hline
  repr  &   dim  &  $T_{F}(r)$  &  $[SU(5)]^3$  & $U(1)_{\psi\chi}\,[SU(5)]^2$  &   $[U(1)_{\psi\chi}]^3$  \\
 \hline
  ${\small \yng(2)} \phantom{\bar{\small \yng(1)}} \! \! \! \! \! \! $     
& $21$   &  $0$   &  $0$    &  $0$  &  $2625$ \\
    \hline  
     ${\small {\bar {\yng(1,1)}}} \phantom{\frac{1}{A_1} {\small {\bar {\bar {\yng(1,1)}}}} } \! \! \! \! \! \! \! \!  \! \! \! \! \!$    
 & $15\cdot 5$   &  $\frac{1}{2}$   &  $15$  &  $\frac{1}{2}  \cdot (-2) \cdot 15=-15$    & $ -600$ \\
    \hline  
\end{tabular}
 \] 
and so in total:
\bqa 
  A_{ \rm UV}\big([SU(5)]^3\big) &=&  15\ , \nonumber \\ 
   A_{ \rm UV}\big(U(1)_{\psi\chi}\,[SU(5)]^2\big)   &=& \frac12\cdot(-2)\cdot 15=-15 \ , \nonumber \\
   A_{ \rm UV}\big([U(1)_{\psi\chi}]^3\big)  &=& 21 \cdot5^3 - 15\cdot2^3\cdot5=2025  \ .     \label{anomaliesUV}
\eea

Let us investigate whether the system can confine
without any condensates forming. 
We ask  if color-singlet  massless ``baryon"  states  can be formed which would saturate the above anomalies.  The only (simple) possibility is to 
contract the color as 
\be \bar{\yng(1,1)} \otimes \bar{\yng(1,1)} \otimes \bar{\yng(1,1)}   =  (\cdot) + \ldots  \ee
that is, 
\be B^{\gamma, I,J, K} = \epsilon_{ijlmno}\chi^{\alpha, I}_{ij}\chi^{\beta, J}_{lm}\chi^{\gamma, K}_{no} \epsilon_{\alpha\beta}
\ee
where Greek letters are spin, small Latin are color, big Latin are flavor.
A priori these can belong to  different $SU(5)$ flavor  representations,
\be \yng(1) \otimes  \yng(1) \otimes  \yng(1) =\yng(1,1,1)+\yng(2,1)+\yng(2,1)+\yng(3)\ . \ee
Actually the first (completely antisymmetric) and the last (completely symmetric)  are  both excluded by the statistics. 
We are left with a mixed representation ${\tiny  \yng(2,1)} $.
Its anomaly content is 
\[
  \begin{tabular}{|c|c|c|c|c|c|}
\hline
  repr  &   dim  &  $T_F(r)$  &  $[SU(5)]^3$  & $U(1)_{\psi\chi}\,[SU(5)]^2$  &   $[U(1)_{\psi\chi}]^3$  \\
 \hline
  ${\small \yng(2,1)} \phantom{\bar {\small \yng(1,1)}} \! \! \! \! \! \! \!  $     & $40$   &  $11$   &  $16$  &  $ 11\cdot  (-6)=-66$  &  $40\cdot   (-6)^3=-8640$ \\
    \hline  
\end{tabular}
\]
Clearly these baryons cannot reproduce the  anomalies  (\ref{anomaliesUV})  due to the fermions in the UV theory. 
We conclude that if the system confines in the IR, with a gapped vacuum (vacua), the global symmetry (\ref{global}) must be broken spontaneously, at least partially.

~~~

Let us check whether  the 1-form gauging of a center symmetry can give any useful information. 
In our system the surviving exact center symmetry is ${\mathbbm Z}_{2}^{C} \subset {\mathbbm Z}_{6}^{C} $: 
\be     {\mathbbm Z}_{2}^{C}\,:\quad     e^{ i \oint A } \to    e^{ \frac{2\pi i}{ 2}  k }      \,   e^{ i \oint A }   \ ,  \qquad        k=1,2\ .
\ee 
We gauge this 1-form symmetry and study its effects on the discrete symmetries,  by introducing the fields
\begin{itemize}
\item $A_{\psi}$: $ {\mathbbm Z}_{8}^{\psi}$ 1-form gauge field,
\item $A_{\chi}$: $ {\mathbbm Z}_{20}^{\chi}$ 1-form gauge field,
\item $B^{(2)}_c$: $\mathbb{Z}_{2}^C$ 2-form gauge field.   
\end{itemize}
The last satisfies 
\be    2 B_c^{(2)} = d B_c^{(1)}\ , \qquad   B_c^{(1)}  \to  B_c^{(1)} + 2 \lambda\ ,  \qquad  B_c^{(2)} \to B_c^{(2)}  + d \lambda   
\ee
The $6$D anomaly functional is
\bea   &&    \frac{1}{24 \pi^2}           {\tr}_{\tiny \yng(2)}  \big(  {\tilde F} -  B_c^{(2)}      -    d   A_{\psi} \big)^3  +       \frac{1}{24 \pi^2}         {\tr}_{\tiny  {\bar {\yng(1,1)}}}  \big(   {\tilde F} -  B_c^{(2)}       -    d   A_{\chi}\big)^3   \nonumber \\
&&=    -    \frac{8}{8\pi^2}       {\tr}_{} \big(   {\tilde F} -  B_c^{(2)} \big)^2  \wedge d A_{\psi}   -     \frac{20}{8\pi^2}       {\tr}_{}  \big(   {\tilde F} -  B_c^{(2)} \big)^2  \wedge d A_{\chi}   + \dots \ .   \eea
The relevant part of the $5$D  WZW action is
\be   -   \frac{8}{8\pi^2}     \left[   \int   {\tr}_{}  {\tilde F}^2  -  6   \, (B_c^{(2)} )^2   \right]    \, A_{\psi}  
   -     \frac{20}{8\pi^2}    \left[   \int   {\tr}_{}  {\tilde F}^2  -  6   \, (B_c^{(2)} )^2   \right]   \, A_{\chi}  \ .   \label{5DWZW}
  \ee
  ${\mathbbm Z}_8^{\psi} $  and  ${\mathbbm Z}_{20}^{\chi}$ transformations are expressed by the variations
  \be     A_{\psi} = d A_{\psi}^{(0)}\;, \quad  \delta A_{\psi}^{(0)} =  \frac{2\pi k }{8}\;;   \qquad 
      A_{\chi} = d A_{\chi}^{(0)}\;, \quad  \delta A_{\chi}^{(0)} = -  \frac{2\pi \ell }{20}\;,    \label{Z8Z20}
  \ee
  ($ k \in {\mathbbm Z}_8$, $ \ell \in {\mathbbm Z}_{20}$).
The integration over closed $4$ cycles give 
\be     \frac{1}{8\pi^2}    \int   {\tr}_{}  {\tilde F}^2   \in {\mathbbm Z}\ , \qquad   \frac{1}{8\pi^2}    \int    (B_c^{(2)} )^2    \in  \frac{  {\mathbbm Z}}{4}\ .
\ee
Therefore
one finds that both ${\mathbbm Z}_8^{\psi} $  and  ${\mathbbm Z}_{20}^{\chi}$
are broken by the 1-form ${\mathbbm Z}_2^{C} $ gauging: 
\be      {\mathbbm Z}_8^{\psi} \longrightarrow    {\mathbbm Z}_4^{\psi}\ , \qquad  {\mathbbm Z}_{20}^{\chi} \longrightarrow    {\mathbbm Z}_{10}^{\chi} \ .\label{suggests}
\ee
\be  \delta A_{\psi}^{(0)} =  \frac{2\pi k}{8}\ , \quad k=2,4,\ldots, 8\ ,  \qquad   \delta A_{\chi}^{(0)} = - \frac{2\pi \ell}{20}\ , \quad \ell=2,4,\ldots, 20\ .
\ee
Such a result suggests a nonvanishing condensate of some sort to form in the infrared, and breaks the global symmetry at least partially.

Actually, a more careful analysis is needed to see which bifemion condensates may occur in the infrared, in order to satisfy the mixed-anomaly-matching condition. 
The division by  ${\mathbbm Z}_{40}$ in the global symmetry group, (\ref{global}),  is due to the fact that the subgroup (\ref{Z40}) 
is inside the nonanomalous $U_{\psi\chi}(1)$. The quotient
\be    {\mathbbm Z}_4 \sim  \frac{  {\mathbbm Z}_{20} \times  {\mathbbm Z}_8}{{\mathbbm Z}_{40}}   \label{quotient}
\ee 
also forms a subgroup, which can be taken as 
\be  \psi \rightarrow e^{2\pi i \frac{2k}{8}}\psi=e^{2\pi i \frac{k}{4}}\psi \;; \qquad   
\chi \rightarrow e^{-2\pi i \frac{5k}{20}}\chi=e^{-2\pi i \frac{k}{4}}\chi\;,   \label{Z40Bis}
\ee
or  
as
\be  \delta A_{\psi}^{(0)} =  \frac{2\pi k }{4}\ , \qquad   \delta A_{\chi}^{(0)} = - \frac{2\pi k }{4}\ , \qquad  k=1,2,3, 4\,,   \label{fromZ4}
\ee
in  (\ref{5DWZW}), (\ref{Z8Z20}).

The action of ${\mathbbm Z}_{40}$ on the $4D$ partition function can be obtained by setting 
$k=\ell=1,2,\ldots, 40$, in  Eq.~(\ref{Z8Z20}),  or the chiral transformations,  Eq.~(\ref{Z40Bis}).  The anomaly is proportional to 
\be   -   \left(  8  \cdot   \frac{2\pi k }{8} -  20\cdot  \frac{2\pi k }{20} \right)  \,  
\frac{1}{8 \pi^2}    \left[   \int   {\tr}_{}  {\tilde F}^2  -  6   \, (B_c^{(2)} )^2   \right]   =0\;:
\ee
i.e.,  ${\mathbbm Z}_{40}$ remains nonanomalous, even after 1-form gauging of   ${\mathbbm Z}_2^{C} $ is done.

On the other hand,   ${\mathbbm Z}_{4}$   is affected by the gauging of the  center ${\mathbbm Z}_2^{C} $ symmetry.  From  (\ref{fromZ4}) and (\ref{5DWZW})   one finds that the $4D$ anomaly is given by
\be   -  3\cdot  2\pi k \,  
\frac{1}{8 \pi^2}    \left[   \int   {\tr}_{}  {\tilde F}^2  -  6   \, (B_c^{(2)} )^2   \right]   =2\pi k  \cdot \left(   {\mathbbm Z} +    3\cdot 6\cdot  \frac{\mathbbm Z}{4} \right) \;.
\ee
Clearly ${\mathbbm Z}_{4}$   is reduced to   ${\mathbbm Z}_{2}$    ($k=2,4$)  by the 1-form gauging of ${\mathbbm Z}_2^{C} $.

Having learned  the fates of the  discrete symmetries 
\be    {\mathbbm Z}_{20} \times  {\mathbbm Z}_8 \sim    {\mathbbm Z}_{40} \times  {\mathbbm Z}_4
\ee
under the gauged 1-form center symmetry  ${\mathbbm Z}_2^{C} $, let us discuss now what their implications on the possible 
condensate formation in the infrared are. Restricting ourselves to the three types of bifermion condensates, 
\be    \psi \chi\;,  \qquad   \psi \psi,   \qquad \chi \chi\;,  
\ee
the MAC criterion might suggest some condensates in the channels
\bea & & A:  \qquad  \psi \left(\raisebox{-2pt}{\yng(2)}\right) \, \psi \left(\raisebox{-2pt}{\yng(2)}\right)   \quad  {\rm forming}  \quad   \,  \raisebox{-6pt}{\yng(2,2)}\;;
\nonumber \\  & & B:  \qquad  \chi \left(\bar{\raisebox{-9pt}{\yng(1,1)}}\right) \, \chi  \left(\bar{\raisebox{-9pt}{\yng(1,1)}}\right)  \qquad \ \   {\rm forming}  \quad \bar{\raisebox{-12pt}{\yng(1,1,1,1)}}\;;
\nonumber \\  && C:  \qquad   \psi  \left(\raisebox{-2pt}{\yng(2)}\right)   \, \chi \left(\bar{\raisebox{-9pt}{\yng(1,1)}}\right)  \quad \  \ {\rm forming  ~ adjoint ~ representation}\,\, ;\ \label{condensates}
 \eea
 with the one-gluon exchange strengths   ($N=6$) proportional to,
  \bea  & &  A:  \qquad     \frac{2 (N^2-4)}{N} -     \frac{ (N+2)(N-1)}{N}  -     \frac{ (N+2)(N-1)}{N}    = - \frac{2 (N+2)}{N}  \;;
\nonumber \\ & &  B:  \qquad     \frac{2 (N+1)(N-4))}{N} -     \frac{ (N+1)(N-2)}{N}  -     \frac{ (N+1)(N-2)}{N}    = - \frac{4 (N+1)}{N}  \;;
\nonumber\\  & &  C:  \qquad   N-     \frac{ (N+2)(N-1)}{N}  -     \frac{ (N+1)(N-2)}{N}    =   - \frac{N^2-4}{N}  \;,
 \eea
 i.e.,  $ 16/6,  28/6, 32/6 $,  respectively.   
 Among these the last channel is most attractive, and it is tempting to assume that the only condensate in the infrared is 
\be  \brc (\psi \chi)_{adj} \ckt \ne 0\;.  \label{only}
\ee
However,  the mixed anomalies studied above  require that at least two different types of condensates be formed in the infrared.  The breaking of ${\mathbbm Z}_8$ (acting on $\psi$),    ${\mathbbm Z}_{20}$ (acting on $\chi$), and of  ${\mathbbm Z}_4$ (acting on both $\psi$ and $\chi$ but not on the composite $\psi\chi$),  precludes the possibility that only one type of condensate, for instance,  (\ref{only}),   is formed.  Assuming that any two of the condensates (\ref{condensates})) or all of them,  are formed, the discrete symmetries ${\mathbbm Z}_8$, ${\mathbbm Z}_{20}$, and ${\mathbbm Z}_4$ are all broken to  ${\mathbbm Z}_2$ and consistency with the implication of the 1-form gauging of  $\mathbb{Z}_{2}^C$ is attained. 

Actually, we cannot logically exclude the possibility that only one condensate (one of (\ref{condensates})) is formed \footnote{The possibility of symmetric vacuum with no condensate formation has been already excluded.},  part of the color and flavor symmetries survive unbroken, and the associated massless composite fermions
in the infrared might induce, through its own mixed anomalies,  breaking of the remaining unbroken part of discrete symmetries which is "not accounted for" by the unique bifermion condensate.  There are however too many unknown factors in such an argument (which symmetry breaking pattern, which set of massless baryons, etc.), to justify a more detailed discussion on this point.

In any case,  we conclude that the color symmetry is broken at least partially  (dynamical Higgs phase), together with (part of)  flavor symmetry.  Other than this,  the information we
possess at the moment  is unfortunately not powerful enough to indicate in more detail the infrared physics of this system.\footnote{  
Just for completeness,   let us comment also on the conventional  matching condition for the discrete  ${\mathbbm Z}_4$ symmetry, independently of the 
${\mathbbm Z}_4 -  {\mathbbm Z}_2^{C} $ mixed anomalies discussed above.    If  ${\mathbbm Z}_4$ is to remain unbroken  in the infrared,   (\ref{dam})  requires
for $N=4$, that 
$ A_{\rm IR} -  A_{\rm UV}  =  0     \mod   4. $ 
But
 $ A_{\rm UV}\big([\mathbbm{Z}_4^3 ] \big)=21  -  5  \cdot 15  = -  54   \ne  0, \, \mod 4,
 $
 so  a unique confining vacuum with mass gap (no condensates) is not consistent with the conventional  $[{\mathbbm Z}_4]$  anomaly matching conditions
 either.   On the other hand, such a vacuum have been already excluded on the basis of the standard anomaly matching conditions involving $U_{\psi\chi}(1)$ and $SU(5)$.   }

\subsection{$SU(8)$ theory with  \,\underline{$36$}\,$\oplus$\,$3$\,$ \times$\,\underline{${28}$}$^{*}$ }
\label{subscchie}

The classical continuous favor symmetry of this model is
\be
U(1)_{\psi}\times U(1)_{\chi}\times SU(3)\ .
\ee
A nonanomalous $U(1)_{\psi\chi} \subset U(1)_{\psi} \times U(1)_{\chi}$ symmetry has associated charges:
\be 
(Q_{\psi}, Q_{\chi})=(9, -5)\ .
\ee
The system has nonanomalous discrete groups:
\be
U(1)_{\psi}\longrightarrow \mathbb{Z}_{10}^{\psi}\ , \qquad 
U(1)_{\chi}\longrightarrow \mathbb{Z}_{18}^{\chi}\ .
\ee
To check the overlap between $\mathbbm{Z}_{10}\times \mathbbm{Z}_{18}$ and $U(1)_{\psi\chi}$ 
set
\be e^{2\pi i 9\alpha}\psi=e^{\frac{2\pi i }{10}m}\psi\ ,  \qquad  e^{-2\pi i 5\alpha}\chi= e^{\frac{2\pi i }{18}n}\chi\ . \ee
If we write $\alpha = \frac{k}{90}$ we get $k = m + 10 \ell$ and $-k = n + 18 m $  ($k,\ell, m \in {\mathbbm Z}$).
This has solution  for each $m+n$ even.   This means that if $m + n$ is even  $k$ can be chosen such that the $U_{\psi\chi}$ transformation cancel the discrete transformation.
If instead $m + n$ is odd  $k$ can be chosen to erase only one of them, e.g.,  choose $k$ to cancel m and take $n-k$ to be 9.
So the anomaly free symmetry group is:

\be 
SU(3)\times U(1)_{\psi\chi}\times \mathbbm{Z}_2  \label{symmetries}
\ee
where $\mathbbm{Z}_2$ act as:
\bea
\begin{array}{c}
\psi \rightarrow -\psi \\
\chi \rightarrow \chi  
\end{array}
\qquad 
{\rm or}
\qquad 
\begin{array}{c}
\psi \rightarrow \psi \\
\chi \rightarrow -\chi  
\end{array}
\eea
the two representation are equivalent because a $U(1)_{\psi\chi}$ transformation takes the former to the latter.

The anomaly coefficients in the UV are:
\bea A_{\rm UV} \big([SU(3)]^3\big) 
&=& 28\ ,  \nonumber  \\
A_{\rm UV} \big(U(1)_{\psi\chi}\,[SU(3)]^2\big)&=&\frac12 \cdot(-5)\cdot  28=-140\ , \nonumber  \\
  A_{\rm UV} \big([U(1)_{\psi\chi}]^3\big)&=&36 \cdot9^3 - 28 \cdot3\cdot5^3=15744\ . \eea
In our system no gauge invariant spin $\tfrac{1}{2}$  baryons can be formed by using three fundamental fermions. 
Barring the possibilities that some massless baryons made of more than three fermion components,
perhaps with gauge fields,  saturate such anomalies,   one is forced to conclude that 
$SU(3)\times U(1)_{\psi\chi}$ symmetry is broken in the infrared.

Let us first check the 1-form symmetry: this system has an exact  $\mathbb{Z}_2^{C}$ center symmetry.  As usual we can introduce:
\begin{itemize}
\item $A_{\psi}$: $ {\mathbbm Z}_{10}^{\psi}$ 1-form gauge field,
\item $A_{\chi}$: $ {\mathbbm Z}_{18}^{\chi}$ 1-form gauge field,
\item $B^{(2)}_c$: $\mathbb{Z}_{2}^{C}
$ 2-form gauge field.
\end{itemize}
where again:
\be    2 B_c^{(2)} = d B_c^{(1)}\ , \qquad   B_c^{(1)}  \to  B_c^{(1)} + 2 \lambda\ ,  \qquad  B_c^{(2)} \to B_c^{(2)}  + d \lambda   
\ee
The anomaly polynomial is:
\bea   &&    \frac{1}{24 \pi^2}           {\tr}_{\tiny \yng(2)} \big(   {\tilde F} -  B_c^{(2)}      -    d   A_{\psi}\big)^3   +       \frac{1}{24 \pi^2}        {\tr}_{\tiny  {\bar {\yng(1,1)}}} \big(  {\tilde F} -  B_c^{(2)}       -    d   A_{\chi}\big)^3   \nonumber \\
&&=    -    \frac{10}{8\pi^2}        {\tr}_{} \big(   {\tilde F} -  B_c^{(2)} \big)^2  \wedge d A_{\psi}   -     \frac{18}{8\pi^2}         {\tr}_{}  \big(   {\tilde F} -  B_c^{(2)} \big)^2  \wedge d A_{\chi}   + \dots \ .   \eea
The relevant part of the $5$D  WZW action is therefore
\be   -   \frac{10}{8\pi^2}     \left[   \int   {\tr}_{}  {\tilde F}^2  -  8   \, (B_c^{(2)} )^2   \right]    \, A_{\psi}  
   -     \frac{18}{8\pi^2}    \left[   \int   {\tr}_{}  {\tilde F}^2  -  8   \, (B_c^{(2)} )^2   \right]   \, A_{\chi}  \ .
  \ee
The integration over closed $4$ cycles give this time
\be     \frac{1}{8\pi^2}    \int   {\tr}_{}  {\tilde F}^2  \in {\mathbbm Z}\ , \qquad   \frac{1}{8\pi^2}    \int    (B_c^{(2)} )^2    \in \frac{  {\mathbbm Z}}{4}\ .
\ee
Note that, in contrast to all other cases studied in this paper  (except for the $SU(4)$ model in Section~\ref{sec:special}),  the 1-form gauging of the exact $\mathbb{Z}_2^{C}$ center symmetry this time 
does not lead to the mixed anomalies,  i.e., does not imply breaking of  the discrete ${\mathbbm Z}_{10}^{\psi}$ or ${\mathbbm Z}_{18}^{\chi}$  symmetries. It does not give any new information on the infrared dynamics.

Because of the difficulties in satisfying the conventional 't Hooft anomaly constraints,  one is led to believe that the symmetry of the model
(\ref{symmetries})  is spontaneously broken in the infrared, by some condensate.   It is possible that the vacuum (or vacua) is (are) characterized by four-fermion
condensates, but the simplest scenario seems to be  dynamical Abelianization, triggered by a bi-fermion condensate,
\be  \brc \psi \chi \ckt   
\ee
in color contraction
 \be      \yng(2) \otimes  {\bar {\yng(1,1)}} =   \yng(2,1,1,1,1,1,1) + \ldots\ , 
  \ee
i.e.,     in the adjoint representation of color $SU(8)$.   The global symmetries could be  broken as
\be  
SU(3)\times U(1)_{\psi\chi} \times \mathbbm{Z}_2  \longrightarrow  SU(2)\times U(1)_{\psi\chi}^{\prime}\ , 
\ee
where $U(1)_{\psi\chi}^{\prime}$ is an unbroken combination of $SU(3)$ and $U(1)_{\psi\chi}$.  In this case the system may dynamically Abelianize completely \cite{BKS,BK}. 
The conventional 't Hooft anomaly conditions are satisfied by the fermion components which do not condense and remain massless, in a simple manner, as in the 
model considered in the previous subsection.

\section{Discussion}
\label{conc}

In this paper symmetries and dynamics of several  gauge theories which possess an exact 
center symmetry have been studied. Let us consider a general setup which includes all the cases studied here. We consider an  $SU(N)$ gauge theory with matter content consisting of Weyl fermions $\psi_i$ in representations $R_i$ and multiplicities $N_{f,i}$ with $i=1, \dots, n_R$ where $n_R$ is the number of different representations (in the cases discussed $n_R$ has been at most 2). We consider only cases in which the gauge anomaly cancels and  where $b_0$ is positive
(asymptotically free theories). Each $U(1)_{\psi_i}$ global symmetry is broken due to instantons  as
\be U(1)_{\psi_i} \longrightarrow {\mathbbm Z}_{2 T_{R_i} N_{f,i}}^{\psi} \ .
\ee

Every representation of $SU(N)$ has a certain $N$-ality associated to it; that is the way $\psi_i$ transforms under the center of the gauge group ${\mathbbm Z}^C_N$ and  corresponds  to the number of boxes in the Young tableaux modulo $N$. Let  $n(R_i)$ be the $N$-ality of the representation $R_i$.
 For example the fundamental representation has $N$-ality $1$, the two-index representations  have $N$-ality $2$ and the adjoint has $N$-ality $0$.    We then consider the greatest common divisor between $N$ and all the $N$-alities of the various representations 
\beq
k = \gcd{\big(N,n(R_1),n(R_2),\dots,n(R_{n_R})\big)} \ .
\eeq
If $k$ is greater than 1, then we have a nontrivial $1$-form center symmetry ${\mathbbm Z}^C_{k}$. 
A toron can be constructed with $n_{12}=n_{34}=\frac Nk $ units of 't Hooft fluxes and it has topological  charge equal to 
\be    \frac{n_{12} \, n_{34} }{N}      =   \frac{1}{N}   \frac{N^2}{k^2} =   \frac{N}{k^2} \ee
 that of the instanton. This can be combined with some integer (instanton) number  to yield  the minimal possible topological charge 
\beq
\frac{1}{\tk} =
\frac{\gcd{\big(\frac{N^2}{k^2},N\big)}}{N} \ .
\eeq

To see this,    set
\be    N =  k \,n\;,  \qquad  n  \in {\mathbbm Z}\;.
\ee
then
\be {\tilde k}=   \frac{N}{\gcd{\big(\frac{N^2}{k^2},N\big)}} =  \frac{k n}{\gcd{\big(  n^2, k n\big)}}=      \frac{k}{\gcd{\big(  n, k \big)}} \in {\mathbbm Z}\;.\ee
Now the toron charge is
\be    \frac{N}{k^2}=   \frac{n}{k}  =   \frac{n/\gcd(n,k)}{k/\gcd(n,k)} =    \frac{m }{k/\gcd(n,k)}   = \frac{m}{\tilde k}\;, \qquad   m \equiv  \frac{n}{\gcd(n,k) } \in {\mathbbm Z}\,, 
\ee
where $m$ and ${\tilde k}$ are coprime.  
Combining this with some instanton number $q$,     one has
\be    \exists p,  \exists  q \in {\mathbbm Z}\;, \qquad 
   \frac{N}{k^2}  \cdot p +   q   =    \frac{m }{\tilde k}  \cdot p + q =   \frac{  m p + {\tilde k} q}{   {\tilde k}}  =  \frac{1}{{\tilde k}}   \;,  \ee
due to B\'ezout's lemma.

If this is a fractional number, i.e. if $\tk $ is an integer larger than 1,    we have generalized (mixed) anomalies of the type ${\mathbbm Z}_{2 T_{R_i} N_{f,i}}^{\psi}[{\mathbbm Z}^C_{k}]^2$,   and the discrete symmetry is  further broken as
\be
 {\mathbbm Z}_{2 T_{R_i} N_{f,i}}^{\psi} \longrightarrow    {\mathbbm Z}_{2 T_{R_i} \! N_{f,i} / {\tilde k}   }
\ee
That ${\tilde k}$ is a divisor of  $2 T_{R}$ can be shown, case by case,  by using the formulas given in Appendix~\ref{Dynkin},  as  has been verified in all cases encountered. 

Note that the existence of a nontrivial  center symmetry does not necessarily imply the presence of generalized (mixed) anomalies, as $k$ and $\tk$ may be different. We have seen three examples in this paper, the $SU(4)$ model in Subsection \ref{sec:special},  $SU(4\ell)$  cases in Section~\ref{two},  and the $SU(8)$ chiral  model in Subsection \ref{subscchie},
where  ${\tilde k}=1$ and where no mixed anomalies arise.

In all  cases studied in the paper the presence of some  bi-fermion condensates would explain the anomaly matching by breaking the discrete chiral symmetry down to a sufficiently small subgroup. Most of the time this subgroup is smaller than the minimal one required, e.g., from matching of the mixed anomalies. In any event,  the use of  mixed and conventional  't Hooft anomaly matching constraints  in general provides us with  significant, if not  decisive, indications about the infrared dynamics of the theories.

\section*{Acknowledgments}
 We thank  Marco Costa and Yuya Tanizaki  for useful discussions. This work  is  supported by the INFN special project grant ``GAST (Gauge and String Theories)".

\appendix
\section{The Dynkin index of some  $SU(N)$ representations \label{Dynkin} }
 The Dynkin index $T_R$ is defined by
\be   {\rm {tr}} \big( t_R^a t_R^b \big) =  T_R  \,  \delta^{ab}\,,
\ee
where $t_R^a$  are the generators of $SU(N)$ in the representation $R$. 
Summing over $a=b$, one gets
\be    d(R)  C_2(R) =   T_R \, (N^2-1)\ ,  \qquad     \sum_a t_R^a  t_R^a = C_2(R) {\mathbbm 1}_{d(R)}\ ,
\ee
where $d(R)$ is the dimension of the representation and $C_2(R)$ is the quadratic Casimir. 
For the fundamental representation one has
\be      C_2(F) = \frac{N^2-1}{2N}\ ,  \qquad d(F)=N\ , \qquad   T_{F} =\frac{1}{2}\ ,
\ee
and for the adjoint, 
\be     C_2({\rm adj}) = N\ ,  \qquad d({\rm adj})=N^2-1\ , \qquad   T_{\rm adj} = N\ ;
\ee
these two are quite familiar. 
For a rectangular Young tableau the quadratic Casimir is\footnote{See for example the book of   Barut and Raczka \cite{Barut},  p. 259 (apart from a factor $1/2$ which is included, so that  $C_2(F) = (N^2-1)/2N$ for the fundamental). See also \cite{Walking} for reference.}
 \be   C_2 (  \overbrace{f,  \dots,  f }^k, 0,\dots) =  \frac{k f (N+f) (N-k)}{2 N}\ ,
 \ee 
 where $f$ is the number of the boxes in a row and $k$ the number of rows.

For the order $n$-antisymmetric representation,  $f=1$, $k=n$,      it  is 
\be   C_2(R) = \frac{n (N-n) (N+1)}{2N}\ .
\ee
Taking into account the multiplicity
 \be d(R)=\frac{N(N-1) \cdots (N-n+1)}{n!}=  \frac{N!}{n!  (N-n)!}   \ , 
 \ee
 the Dynkin index of totally antisymmetric single column  representation of height $n$   is given by
 \be    T_R =  \frac{ (N-2)(N-3) \cdots (N-n)}{2 (n-1)!}\ .
\ee

For the special cases  (of relevance in Section~\ref{sa})  with  $N= 2n$ even,  we have
\be     C_2(R) =   \frac{N(N+1)}{8}\ ,     \qquad   d(R)=      {N \choose  N/2}\ .
\ee
and
\be    2\, T_R =     {N-2 \choose  N/2-1}=   {2n -2 \choose  n-1} \ .\label{thisexp}
\ee
By using this expression    it is easy to see  that  for    $N= 4 \ell$,    $n= 2\ell$,  $2\, T_R$ is a multiple of $ \ell $,  whereas for 
 $N= 4 \ell+2$,   $2\, T_R$  contains $2\ell+1 $ as a divisor. To prove it,  note that the general  combinatoric number
 \be   
   {m \choose  r}=  \frac{m (m-1)\ldots (m-r+1)}{r!} =\frac{m!}{r! (m-r)!}\ , \label{toprove1}
 \ee
 is always an integer.  But  
 \be  {m \choose  r}=  {m \choose  r-1} \cdot \frac{m-r+1}{r}\; 
 \ee
 and both  $ {m \choose  r}$ and $ {m \choose  r-1}$ are integers. 
Therefore  $m-r+1$ is a divisor of  $ {m \choose  r}$.  Applying this for  $N= 4\ell$ one finds that 
\be    2\, T_R =     {4 \ell -2 \choose  2\ell  -1} 
\ee
has a divisor  $2 \ell$ hence $\ell$. 
 For $N=4\ell+2$,  
 \be    2\, T_R =     {4 \ell  \choose  2\ell }    \label{toprove2}
\ee
has a divisor,    $2\ell + 1$.

For the symmetric representation of rank $2$, $f=2$, $k=1$, so 
\be    C_2(R)= \frac{ (N+2) (N-1)}{N}\ .
\ee
By taking into account the multiplicity,
\be    d(R)=\frac{N(N+1)}{2} \ , 
 \ee
one finds
\be   2  \,T_R=    2  \,   \frac{ d(R)  C_2(R) }{N^2-1}=  N+2\, .
\ee

For the symmetric representation of rank $m$, $f=m$, $k=1$, so 
\be    C_2(R)= \frac{m  (N+m) (N-1)}{
2 N}\ .
\ee
By taking into account the multiplicity,
\be    d(R)=\frac{(N+m-1)!}{m!  \, (N-1)!} \ , 
 \ee
one finds
\be  2\, T_R=      \frac{ d(R)  C_2(R) }{N^2-1}=\frac{(N+m)!}{(N+1)!\, (m-1)!}\ .
\ee
$2\,T_R$  is a multiple of $m$, as can be shown following a similar consideration as  (\ref{toprove1})-(\ref{toprove2}).

\end{document}